\documentclass[%
 reprint,
 amsmath,amssymb,
 aps,
]{revtex4-2}

\usepackage{graphicx}% Include figure files
\usepackage{dcolumn}% Align table columns on decimal point
\usepackage{bm}% bold math
\usepackage{xcolor}
\usepackage{blindtext}
\usepackage{mathptmx}
\usepackage{etoolbox}
\usepackage{mathrsfs}
\usepackage{babel}
\usepackage{algpseudocode}

\begin{document}

\preprint{AIP/123-QED}

\title{Harnessing Plasmonic Interference for Nanoscale Ultrafast Electron Sources}

\author{Alimohammed Kachwala}
\email{Authors to whom correspondance should be addressed: akachwal@asu.edu}
\author{Mansoure Moeini Rizi}
\affiliation{Department of Physics, Arizona State University, Tempe, AZ 85287, USA}
\author{Christopher M Pierce}
\affiliation{Department of Physics,University of Chicago, Chicago, IL 60637, USA}
\author{Daniele Filippetto}
\affiliation{Advanced Light Source Accelerator Physics Division, Lawrence Berkeley National Laboratory, Berkeley, CA 94720, USA}
\author{Jared Maxson}
\affiliation{Department of Physics,Cornell University, Ithaca, NY 14850, USA}
\author{Siddharth Karkare}
\email{karkare@asu.edu}
\affiliation{
Department of Physics, Arizona State University, Tempe, AZ 85287, USA}%

\date{\today}% It is always \today, today,
             
\begin{abstract}
In this paper we demonstrate the use of plasmonic focusing in conjunction with non-linear photoemisison to develop geometrically flat nanoscale electron sources with less than 40 pm-rad root mean squared (rms) normalized transverse emittance. Circularly polarized light is incident on a gold Archimedean spiral structure to generate surface-plasmon-polaritons which interfere coherently at the center resulting in a 50 nm rms emisison area. Such a nanostructured flat surface enables simultaneous spatio-temporal confinement of emitted electrons at the nanometer and femtosecond level and can be used as an advanced electron source for high-repetition-rate ultrafast electron diffraction and microscopy experiments as well as next-generation of miniaturized particle accelerators.

%By exploiting plasmonic focusing of light, we achieve a remarkably low root mean square normalized transverse electron emittance of less than 40 pm-rad from a geometrically flat metal photocathode. This is facilitated by focusing light to a nanometer scale using plasmonic Archimedean spiral structure and utilizing non-linear photoemission mechanism, resulting in an electron emission spot size of approximately 50 nm root mean square. Such nanostructured electron source, which enables simultaneous spatio-temporal confinement at the nanometer and femtosecond levels, demonstrates potential as an advanced electron source for high repetition rate ultrafast electron diffraction and microscopy apparatus as well as next-generation of miniaturized accelerator applications.

%In this work we demonstrate the generation of a record low root mean square normalized transverse electron emittance of less than 40 pm-rad from a geometrically flat photocathode. This was achieved by harnessing plasmonic focusing of light to a sub-diffraction regime using plasmonic Archimedean spiral structures resulting in a $\approx$50 nm root mean square electron emission spot. Such a nanostructured electron source, which utilizes plasmonic focusing and exhibits simultaneous spatio-temporal confinement to nanometer and femtosecond level can be used as an advanced electron source for the high repetition rate ultrafast electron diffraction and microscopy apparatus as well as next generation accelerator applications.
\end{abstract}

\maketitle

High repetition rate ($>$100 kHz) sub-picosecond pulsed electron beams are critical to the studying of the ultrafast structural dynamics of atomic lattices as well as molecular species through techniques like stroboscopic ultrafast electron diffraction and microscopy (UED/M) \cite{sood2021universal,siddiqui2023relativistic,durham2020relativistic,gliserin2014towards,ji2019ultrafast,ischenko1983stroboscopical,ischenko1994structural,kirchner2013ultrashort}. Even though field emission tips can generate brighter electron beams resulting in sub-angstrom scale spatial resolutions in electron microscopes, they cannot be switched at sub-microsecond timescales, making femtosecond-laser triggered photoemission of electrons a preferred way of generating such sub-picosecond scale electron bunches \cite{siddiqui2023relativistic,durham2020relativistic,ji2019ultrafast,kirchner2013ultrashort}. 

For UED/M applications, the transverse coherence length, $L_c = \frac{\lambdabar_{c} \sigma_{x,s}}{\epsilon_{n,x}}$ determines the the spatial as well as momentum resolution of the instrument \cite{musumeci2018advances,li2022kiloelectron,filippetto2016design}. Here, $\lambdabar_{c}$ is the reduced Compton wavelength of electron, $\sigma_{x,s}$ is the root mean square (rms) size of the electron beam at the sample and $\epsilon_{n,x}$ is the normalized transverse emittance in the one of the two transverse directions ($x$). 
Thus, to enhance the resolution of UED/M apparatus, it is imperative to increase the transverse coherence length of the electron bunches. Further, obtaining high resolution diffraction patterns from regions of solids having dimensions at the nanometer scale calls for the need of nano-scale electron emitters with sub-nanometer scale normalized transverse emittance of the electron bunch \cite{filippetto2022ultrafast}. 

The normalized transverse emittance of the electron bunch in the one of the two transverse directions ($x$) is expressed in terms of the following equation:
\begin{equation}
\epsilon_{n,x} = \frac{\sqrt{\langle x^2 \rangle \langle p_x^2 \rangle - \langle xp_x \rangle^2}}{m_{e}c} ,
\label{eq:one}  
\end{equation}
where $\sqrt{\langle x^2\rangle}\equiv\sigma_{x}$ is the rms electron spot size and $\sqrt{\langle p_x^2\rangle}\equiv\sigma_{p_{x}}$ is the rms electron momentum spread in the x-direction. $\langle xp_x \rangle$ is the correlation term between the location of emission and the transverse momentum,
${m_e}$ is the mass of an electron and ${c}$ is the speed of light \cite{dowell2009quantum}. 

In addition to the smallest possible emittance UED/M instruments require a high enough current (or number of electrons per bunch) to achieve a good signal to noise ratio and collect data in a reasonable amount of time. Often pinholes can be used to collimate the electron beam to reduce the emittance at the cost of the current \cite{ji2019ultrafast,li2022kiloelectron,feist2017ultrafast,siddiqui2023relativistic}. 

The emittance and current (or electrons/bunch) can be combined into one figure of merit, the 4D-Brightness given by: 
\begin{equation}
\mathrm{B_{4D}} = \frac{Q}{\epsilon_{n,rms}^2} ,
\label{eq:Brightness}  
\end{equation}
where $Q$ denotes the total bunch charge and $\epsilon_{n,rms}$ is the geometric mean value of the emittance along the $x$-and $y$-directions \cite{li2022kiloelectron}.

%The 4D brightness and the bunch length can be combined into one figure of merit, the 5D brightness given by:
%\begin{equation}
%\mathrm{B_{5D}} = \frac{\mathrm{B_{4D}}}{\sigma_t} ,
%\label{eq:Brightness5D}  
%\end{equation}
%Schottky and cold field emitters triggered with a laser pulse used in conjunction with the collimating electron optics in a Transmisison Electron Microscope (TEM) have resulted in emittances as low as 13.5 pm-rad  with a current of 101 fA ($\approx$2 electron/shot), rms electron bunch length of 128 fs resulting in ~200 keV energy pulsed electron beams with a 4D-Brightness of $\approx$900 electrons$/$(nm$^{2}$Sr) \cite{feist2017ultrafast}.

Another important factor that often determines the temporal resolution of UED/M apparatus is the rms length of the electron bunches ($\sigma_t$) \cite{filippetto2022ultrafast,musumeci2018advances}. To achieve the best temporal resolutions and mitigate the effects of electron-electron repulsion in a bunch, many UED/M setups rely on photoemisison from flat, large area (several mm scale) cathodes placed in an accelerating electric field in conjunction with radio frequency (RF) bunching cavities \cite{li2022kiloelectron,chatelain2012ultrafast}. Such large area flat cathodes are also required for UED/M instruments based on RF guns used to obtain mega electron volt (MeV) scale energy electron bunches for reduced jitter owing to their relativistic speeds and larger signal in higher-order diffraction peaks due to the shorter electron wavelength \cite{siddiqui2023relativistic,ji2019ultrafast,weathersby2015mega,carter2016transmission}.

Electron beams from such flat cathodes have been collimated using pinholes \cite{li2022kiloelectron,siddiqui2023relativistic,ji2019ultrafast} to result in an emittance of 120 pm-rad with a current of 100-200 fA ($\approx$1 electron/shot) and rms electron bunch length of $<$100 fs. Further improvements in emittance or brightness in such setups requires improvement of brightness at the cathode. 
%For low-repetition rate or single shot UED applications the electron density and hence $\sigma_x$ is often limited to mirometer-scale by the coulomb repulsion between electrons within the bunch\cite{musumeci2018advances}. In such cases the only way to reduce the emittance is reducing $\sigma_{p_{x}}$. However, 
For high repetition-rate UED/M experiments where only single to few electrons per bunch are enough, emittance can be reduced either by $\sigma_x$ or $\sigma_{p_{x}}$ or both. The rms momentum spread ($\sigma_{p_{x}}$) depends on the cathode materials, its surface and the laser fluence \cite{PhysRevLett.128.114801,galdi2021reduction,saha2022physically,feng2017near,cultrera2016ultra,kachwala2023demonstration,saha2021optical,PhysRevApplied.19.014015,knill2023practical,soomary2021performance,PhysRevAccelBeams.26.093401,karkare2014ultrabright}. The rms momentum spread ($\sigma_{p_{x}}$) as low as 50 eV$/c$ has been achieved by cryo-cooling of a copper photocathode  with an atomically ordered surface and operating it at the photoemission threshold \cite{karkare2020ultracold}. The rms electron spot size ($\sigma_x$) is limited by the diffraction limit of light and the ability to focus the laser to a small spot size \cite{silfies2019diffraction,musumeci2018advances}. $\sigma_x$ as small as 1 $\mu$m  has been achieved by operating the cathode in the transmission mode geometry and placing the final focusing lens very close $\sim$1 cm behind the cathode resulting in an rms normalized transverse emittance of about 250 pm-rad \cite{li2022kiloelectron,li2012nanometer,wen2009ultrafast}. At the cathode, the smallest emittance that can be achieved is limited by the Heisenberg's uncertainty principle to $\hbar/2m_ec = $ 0.2 pm-rad. In order to approach this quantum limit, we need nanoscale electron emission area.

Nanoscale electron emisison areas can be achieved by using nanostructures that focus light plasmonically
\cite{gramotnev2010plasmonics,pierce2022towards,durham2019plasmonic}. In such structures, surface-plasmon-polaritons (SPP) excited at the metal-dielectric interface by the incident laser, interfere constructively enhancing the optical field intensity in localized areas on the surface of the metal \cite{durham2019plasmonic,kachwala:ipac2023-tupl188}. Plasmonic Archimedean spiral is one such structure suitable for low emittance, ultrafast nanoscale photoemission. Femtosecond SPP pulses are resonantly excited at the groves of the spiral using a circularly polarized pulsed femtosecond laser. By selecting opposite helicities for the incident circularly polarized light and the spiral, plasmonic focusing at the spiral center can be optimized, achieving nanoscale confinement of the optical field intensity. For instance, compensating the ASP helicity of L = 1 with circularly polarized light of spin angular momentum S = -1 results in an SPP pulse with zero orbital angular momentum (OAM) J = L + S = 0 at the spiral center \cite{guo2017review}. 
At the center of the spiral the electric field is dominated by its out-of-plane component given by a zeroth-order Bessel function $E_{z}(r)\propto J_{0}(k_{spp}r)$, where $k_{spp} = 2\pi/\lambda_{spp}$ and $\lambda_{spp} = $ 783 nm, while the magnetic field is in the azimuthal direction creating a magnetic vortex \cite{guo2017review,gramotnev2010plasmonics,tan2017plasmonic}.

%The resulting electric field at the ASP center is dominated by its out-of-plane component, following a zeroth-order Bessel function $E_{z}(r)\propto J_{0}(k_{spp}r)$, where $k_{spp} = 2\pi/\lambda_{spp}$ and $\lambda_{spp} = $ 783 nm \cite{guo2017review,gramotnev2010plasmonics,tan2017plasmonic}.

In this work we use a plasmonic Archimedean spiral photocathode (ASP) to focus light to nanometer scales  \cite{chen2010experimental} and use non-linear photoemisison to demonstrate an emission spot
($\sigma_x$) of $\approx$50 nm rms resulting in an emittance of less than 40 pm-rad - nearly an order of magnitude smaller compared to the best emittance previously demonstrated from a geometrically flat photocathode \cite{li2012nanometer,jin2008super,maxson2017direct}. The ASP consists of a single groove that completes nine revolutions around the spiral center with the central radius ($R_0$) of 12.5 $\mu$m and the helecity of L = 1. Femtosecond SPP pulses were resonantly excited at the groves of the ASP using circularly polarized (S = -1) femtosecond laser pulses with central excitation energy of 1.55 eV ($\lambda = $ 800 nm). The SPPs propagate to the center of the ASP where they interfere constructively, resulting in the nanoscale electron emission area.

\begin{figure}
\centering
\includegraphics*[width=0.9\columnwidth]{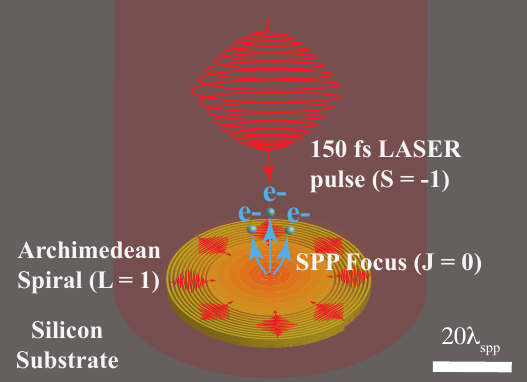}
\caption{Schematic of a gold Archimedean spiral photocathode with L = 1 helicity. Illuminated by a circularly polarized laser with S = -1 spin angular momentum, pulse length of 150 fs, and central wavelength of 800 nm, it produces an SPP pulse with J = 0 orbital angular momentum. The laser is incident at 4° to the gold spiral's surface normal. Constructive interference of generated plasmons at the spiral center enhances intensity. Here, $\mathrm{\lambda_{spp} = }$ 783 nm.}
\label{fgr:experimentalSetup}
\end{figure}
\begin{figure}
\includegraphics*[width=1.0\columnwidth]{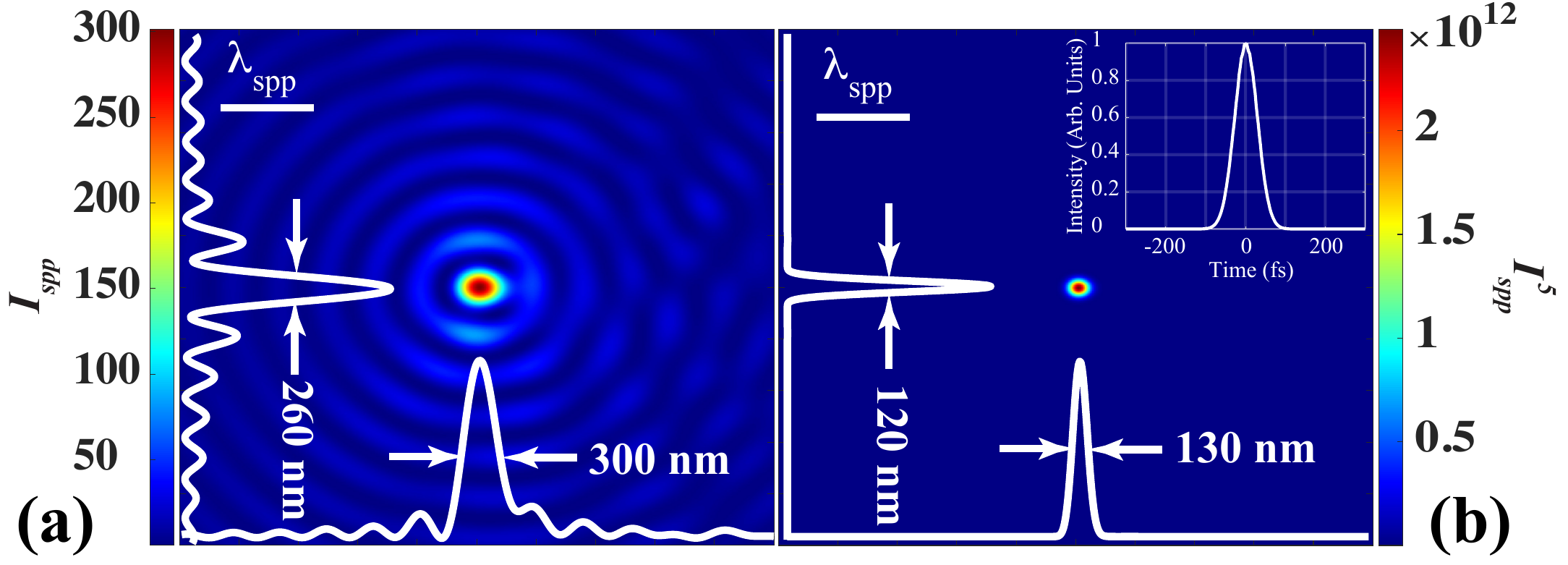}
\caption{Finite difference time domain (FDTD) simulation depicting (a) surface plasmon polariton intensity enhancement $I_{spp}(r) = I(r)/I_{0}(r)$, where $I(r)$ is the measured in-plane intensity and $I_{0}(r)$ is the maximum intensity of the excitation pulse and (b) $I_{spp}^5(r)$ enhancement at the gold ASP center using a circularly polarized Gaussian pulse at 4° incidence angle, central wavelength of 800 nm, and pulse duration of 150 fs as the excitation source. The inset illustrates the ASP's temporal response for $I_{spp}^5(t)$. Here, $\mathrm{\lambda_{spp} =}$ 783 nm.}
\label{fgr:LumericalFDTD}
\end{figure}

The isolated and azimuthally symmetric nano-focus at the center of the spiral has a theoretical full width at half maximum (FWHM) which is smaller than half the SPP wavelength at the metal-vacuum interface \cite{frank2017short}. Figure \ref{fgr:LumericalFDTD} (a) shows the intensity enhancement due to constructive interference of the SPP's following the zeroth order Bessel function at the center of ASP calculated using finite difference time domain (FDTD) simulation using a commercial software suite (Lumerical) \cite{Lumerical} (see Supplementary Information). The spatial extent of the SPP focused intensity [$I_{spp} (r)$] has full width half maximum (FWHM) of $\approx$260 nm. The $5^{th}$ order non-linearity in the electron emission process is proportional to  $I_{spp}^5(r)$ \cite{musumeci2010multiphoton,ferrini2009non}. This further shrinks the electron emission spot to $\approx$120 nm FWHM or $\sigma_x\approx$50 nm as shown in Fig. \ref{fgr:LumericalFDTD} (b). The inset shows the temporal response of the ASP ($\sigma_t \approx$30 fs) assuming the $5^{th}$ order non-linearity in the electron emission process.

The simulated ASP geometry was fabricated by electron-beam lithography (EBL) (see Supplementary Information) and was transferred after a UHV bake-out at 120 °C for 1 day into a commercially available photoemission electron microscope (PEEM) with a 4° angle of incidence \cite{Focus_PEEM}. 
%The schematic of the experimental set-up is shown in Fig. \ref{fgr:experimentalSetup} (see Supplementary Information). 
Initially, the ASP emitted only 0.001 electrons/shot with $\approx$2.5 kW of peak laser power from a $\approx$150 fs, 500 kHz, 800 nm pulsed laser focused down to $\approx$50 $\mu$m FWHM on the ASP. However, the electron's yield increased by several orders of magnitude in less than 5 minutes with peak laser power in the range of 0.7-2.5 kW incident on the ASP. After this initial `activation' step, the enhanced electron emission stayed without any degradation for several days until the ASP was removed from the PEEM UHV environment of 10$^{-10}$ torr into air. Upon re-insertion into the UHV PEEM after the UHV bake, the ASP required reactivation to get the previously enhanced emission. This indicates that the above laser \lq{activation}\rq process leads to cleaning of adsorbates from the atmosphere that were settled on the Au emission surface.

%Before characterization, the ASP was $activated$ with laser peak pulse power in the range of 700-2500 W. Before $activation$, at 2500 W, the ASP was emitting poorly with $<$0.001 electrons/shot which increased to $\approx$1 electron/shot after $activation$. The $activation$ process leads to cleaning of the ASP from any oxides or residual gases settled on it either during fabrication or exposure to the ambient pressure (see Supplementary Information). The base pressure of the PEEM chamber was in the low $10^{-10}$ torr range during activation and characterization. 
%
%Circularly polarized laser with spin angular momentum S = -1, repetition rate of 500 kHz with a pulse length of 150 fs from a pulsed Optical Parametric Amplifier (Light Conversion Orpheus pumped by Light Conversion Pharos) and central wavelength of 800 nm was made incident onto the sample at 4° angle of incidence with respect to the surface normal of ASP. The laser was focused by a lens down to the spot size of $\approx$80 ${\mu}$m. 
%The ASP consists of a single groove that completes nine revolutions around the ASP center to maximize the SPP excitation efficiency. The starting radius ($R_0$) of ASP was 12.5 $\mu$m and the topological charge was L = 1. This topological charge is compensated by circularly polarized laser (S = -1) to yield an SPP
%pulse with vanishing orbital angular momentum (OAM) J = L + S = 0. 

\begin{figure}
\centering
\includegraphics*[width=1.0\columnwidth]{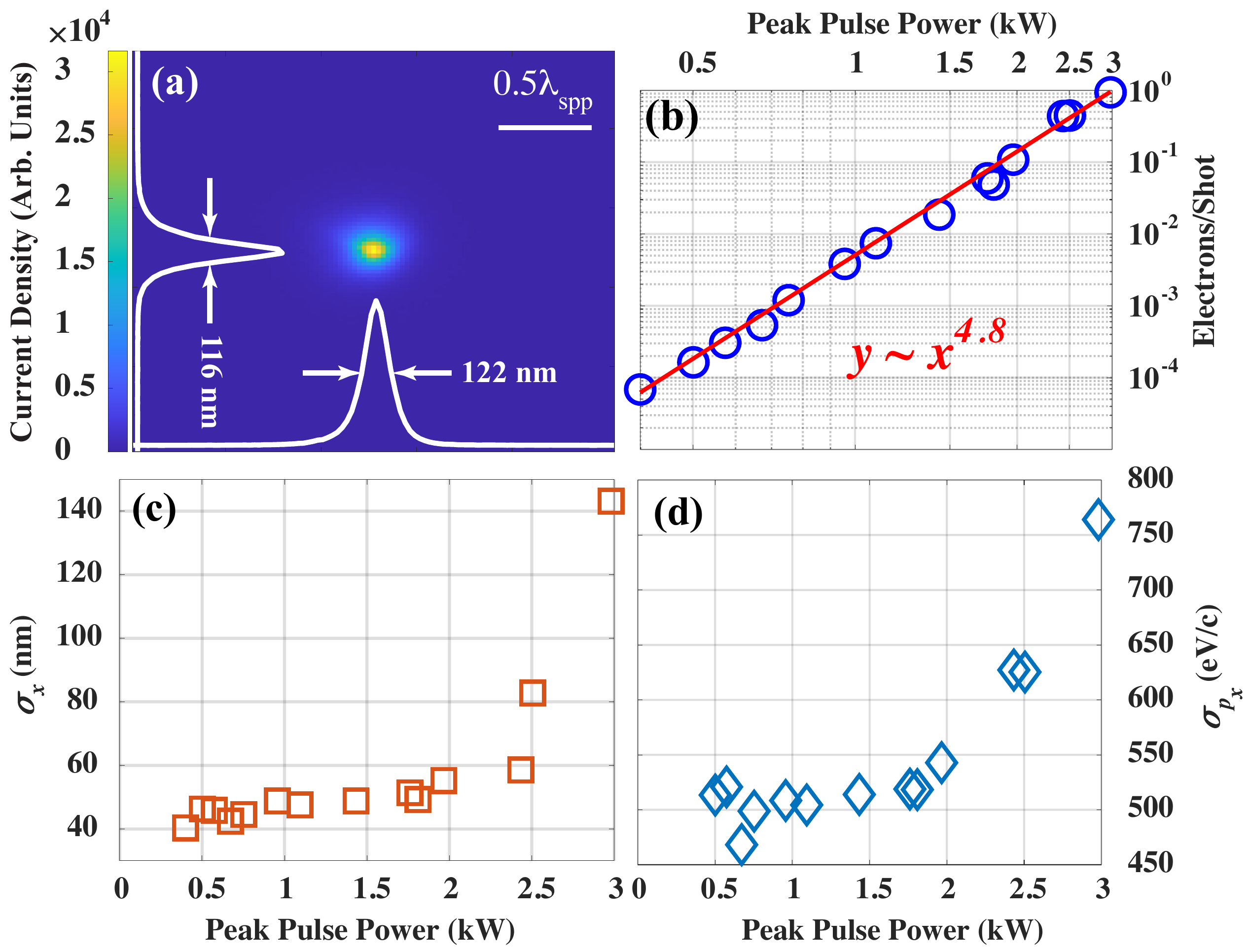}
\caption{(a) Emission profile measured from the center of the ASP using PEEM with incident peak pulse power of $\approx$1.8 kW shown on a linear false color scale. (b) Logarithmic representation of electrons per shot emitted from gold Archimedean spiral photocathode as the function of peak laser pulse power. Evolution of experimentally determined (c) rms electron emission spot size and (d) rms momentum spread in the x-direction as a function of peak pulse power of femtosecond laser with the pulse length of 150 fs and central wavelength of 800 nm. Here $\mathrm{\lambda_{spp} = }$ 783 nm.}
\label{fgr:SpotSize}
\end{figure}

After \lq{activation}\rq, the spatial distribution of non-linear electron emission spot size was measured as shown in Fig. \ref{fgr:SpotSize} (a) at the peak pulse power of $\approx$1.8 kW. As we can see, the rms electron emission spot size is $\approx$50 nm suggesting $5^{th}$ order non-linearity in the photoemission process. The $5^{th}$ order non-linearity is further corroborated by measuring electrons/shot as a function of peak laser pulse power plotted on a double logarithmic scale as shown in Fig. \ref{fgr:SpotSize} (b). Considering the work function ($\phi$) of Au to be $\approx$5.4 eV, at least four quanta of SPP, each with energy $\hbar\omega$ = 1.58 eV are required to overcome the  the work function making 4$^{th}$ order as the lowest possible order of photoemission \cite{pierce2023experimental,pierce2022towards,dreher2023focused}. However, for the case of Au, the large density of $d$-band initial states about 2 eV below the Fermi energy may result in enhanced contribution and thereby lead to enhanced non-linearities in the photoemission process. Inspection of the Au band structure indicates a sharp increase of the joint density of states for total transition energies above 6.4 eV \cite{ramchandani1970energy,rangel2012band,kupratakuln1970relativistic}. At least five quanta of SPP are required to overcome the total transition energies $\approx$6.4 eV. Such $5^{th}$ order non-linearity in the electron emission process has also been observed previously from Au nano-tips in above-threshold photoemission regime \cite{bormann2010tip}.

Figure \ref{fgr:SpotSize} (c) shows evolution of rms electron emission spot size as a function of peak laser pulse power. The rms electron emission spot size is $\approx$50 nm over a wide range of peak pulse power. Figure \ref{fgr:SpotSize} (d) shows evolution of the experimentally determined rms momentum spread of the emitted electrons as a  function of peak laser pulse power. The rms momentum spread of the emitted electrons is $\approx$500 eV/c over a wide range of peak pulse power. 

An average of 0.1 electron/shot are emitted up to a peak pulse power of $\approx$2.5 kW. According to Poisson statistics, this implies $<$1\% probability of 2 electrons being emitted per shot. However, beyond 2.5 kW, 0.5-1 electron/shot is emitted on average, resulting in an 8-20\% probability of emission of 2 electrons/shot. Coulomb interaction between electrons in pulses with 2 or more electrons leads to increased electron emission spot size and rms momentum spread beyond the peak pulse power of $\approx$2.5 kW, as shown in Fig. \ref{fgr:SpotSize} (c) and (d).

%Beyond peak pulse power of $\approx$2.5 kW, on an average, 0.5-1 electron/shot are emitted. Considering Poisson statistics in electron emission with mean of 0.5-1 electron/shot would correspond to 8-20 \% probability in photoemission of 2 electrons/shot. This would result in coulomb interaction between the electrons and thereby result in increase in electron emission spot size as well as MTE beyond the peak pulse power of $\approx$2500 W as seen in Fig.\ref{fgr:SpotSize} (c) and (d).

\begin{figure}
\centering
\includegraphics*[width=1.0\columnwidth]{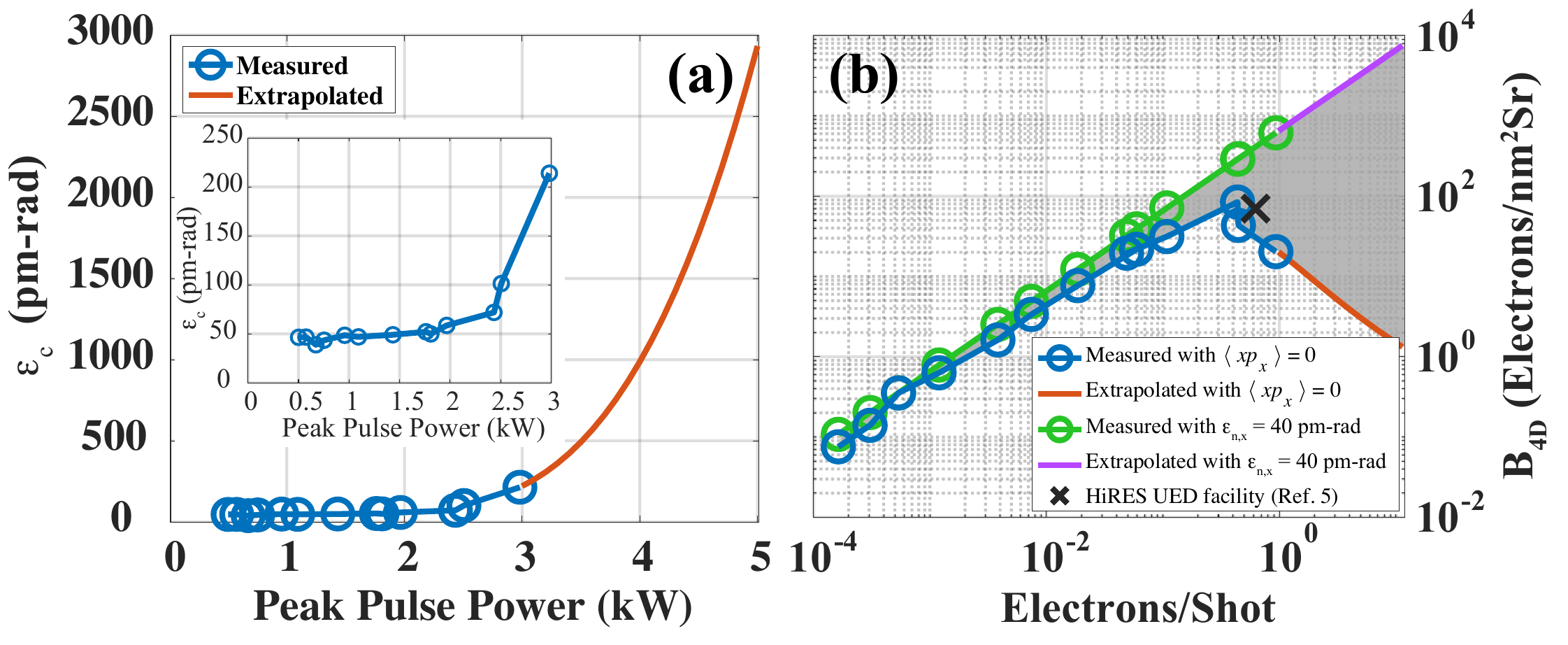}
\caption{(a) Root mean square normalized transverse emittance ($\epsilon_{c}$) is plotted against peak pulse power, where $\epsilon_{c} = \frac{\sqrt{\langle x^2 \rangle \langle p_x^2 \rangle}}{m_{e}c}$. Blue curve: experimental data; red curve: extrapolated data. Inset: zoomed experimental data points. (b) 4D-Brightness of the emitted electron bunch versus electrons/shot. Black cross: 4D-Brightness from Ref. \cite{ji2019ultrafast}. Red and magenta lines: extrapolated 4D-Brightness with $\langle xp_x \rangle = 0$ and $\epsilon_{n,x} = 40$ pm-rad, respectively. Gray shaded region: potential 4D-Brightness range of ASP considering $\langle xp_x \rangle = 0$ or correlated $xp_x$ growth with $\epsilon_{n,x} = 40$ pm-rad.}
\label{fgr:Graphs}
\end{figure}

Ignoring space-phase correlations, i.e. taking $\langle xp_x \rangle = 0$, we calculate the emittance from equation \ref{eq:one}. This is plotted as a function of peak laser pulse power in Fig. \ref{fgr:Graphs} (a). Beyond a peak pulse power of $\approx$2.5 kW, the increase in emittance is attributed to the Coulomb interaction between the emitted electrons. Below $\approx$2 kW with less than 0.1 electron$/$shot the emittance is nearly constant at $\approx$50 pm-rad, with the smallest emittance of 40 pm-rad measured at $\approx$0.001 electron$/$shot. These are the smallest emittances achieved from a geometrically flat photocathode. At 0.5-1 electrons/shot, the normalized transverse emittance measured was in the range of 70-200 pm rad respectively. It was not possible to reliably measure the $\sigma_x$ and $\sigma_{p_{x}}$ beyond 3 kW due to increased Coulomb interactions. Hence we used a cubic fit to extrapolate the emittance beyond these values as shown by the red curve in Fig. \ref{fgr:Graphs} (a).

Schottky emitters triggered with an ultraviolet pulsed laser ($\lambda$ = 400 nm and pulse energy $\approx$10 nJ) have demonstrated a normalized transverse emittance of $\approx$13.5 pm rad at $\approx$2 electrons/shot and $\sigma_t\sim$128 fs after collimation in the TEM gun and column \cite{feist2017ultrafast}. Such tip emitters have been used in electron microscopes, however, face challenges in RF guns due to low lifetime under high fields and emission of dark electrons via field emission asynchronous with the laser pulse that adds background noise to experiments.
The ASP is geometrically flat. Hence these issues related to operation in large electric field RF guns are significantly mitigated. Thus the ASP can enable the use of picometer-scale emittance in RF gun applications like MeV scale UED/UEM.

%It is informative to compare the emittance achieved from the plasmonic ASP to that of the other nano-tip based electron sources. Triggering a Schottky emitter with an ultraviolet pulsed laser ($\lambda$ = 400 nm and pulse energy $\approx$10 nJ), the normalized transverse emittance of $\approx$13.5 pm rad at $\approx$2 electrons/shot and $\sigma_t$$\approx$128 fs has been achieved  after collimation in the TEM gun and column \cite{feist2017ultrafast}.ASP is the only geometrically flat photocathode that has demonstrated pico-meter scale emittance with a potential $\sigma_t$$\approx$30 fs enabling the use of such pico-meter scale emittance sources in high electric field RF guns.

%Such tip emitters typically used in electron microscopes are not easily used in RF guns since under such high fields because of low lifetime. In addition, they tend to emit dark electrons via field emission, which are not synchronous with the laser pulse and add background to the experiment.

%(a) Root mean square normalized transverse emittance with $\langle xp_x \rangle = 0$ plotted against peak pulse power. Here $\epsilon_{c} = \frac{\sqrt{\langle x^2 \rangle \langle p_x^2 \rangle}}{m_{e}c}$. The blue curve shows the experimental data points and the red curve is the extrapolated data. The inset shows the zoomed version of the experimental data points. (b) 4D-Brightness of the emitted electron bunch as a function of electrons/shot. The black cross (\textbf{$\times$}) shows the 4D-Brightness obtained from Ref. \onlinecite{ji2019ultrafast}. The red and magenta lines are the extrapolated 4D-Brightness with $\langle xp_x \rangle = 0$ and $\epsilon_{n,x} = 40$ pm-rad respectively. The gray shaded region is where the 4D-Brightness of ASP can potentially lie considering $\langle xp_x \rangle = 0$ or correlated $xp_x$ growth with $\epsilon_{n,x} = 40$ pm-rad.
%
%In order to characterize ASP in the parameter space of other photocathodes, we compute the 4D-brightness of the electron bunch using equation \ref{eq:Brightness} as shown in Fig.\ref{fgr:Graphs} (b). Assuming $\langle xp_x \rangle = 0$, the maximum 4D-brightness achieved is $\approx$85 electrons$/$(nm$^{2}$Sr). However, considering Coulomb interaction, and thereby correlated $xp_x$ growth, with $\epsilon_{nx} = 40$ pm-rad, the maximum 4D-brightness achieved is $\approx$600 electrons$/$(nm$^{2}$Sr). 
To characterize ASP further,  we compute the 4D-Brightness of the electron bunch using equation 2 and plot it against the electrons per shot as shown in Fig. \ref{fgr:Graphs} (b). 
%Our measurements of $\sigma_x$ and $\sigma_{p_{x}}$ do not give us any information on the underlying correlations between $x$ and $p_x$. 
At the cathode, the correlation term  is $\langle xp_x \rangle = $ 0. For bunch charges with $<$0.1 electron/shot the $\langle x^2\rangle \langle p_x^2\rangle$ is nearly constant (less than 20\% increase) and the brightness increases proportionately with electrons per shot as expected. However, beyond that both $x$ and $p_x$ increases due to Coulomb interactions and $\langle x^2\rangle \langle p_x^2\rangle$ blows up. In this region the brightness calculated by assuming $\langle xp_x\rangle = 0$ reaches a maximum of $\approx$85 electrons$/$(nm$^{2}$Sr) and then reduces due to the increase in $\langle x^2\rangle \langle p_x^2\rangle$. This is shown by the blue curve in Fig. \ref{fgr:Graphs} (b). The red curve corresponds to the brightness obtained from the extrapolated values of $\langle x^2\rangle \langle p_x^2\rangle$ in Fig. \ref{fgr:Graphs} (a). The Coulomb increase in $\langle x^2\rangle$ and $\langle p_x^2\rangle$ also results in increased $x$ and $p_x$ correlations causing $\langle xp_x \rangle$ to be greater than 0. The exact calculations or measurements of the correlations in $x$ and $p_x$ are complex and beyond the scope of this paper. If we assume a fully linearly correlated growth of $x$ and $p_x$, the emittance of 40 pm-rad corresponding to the zero-coulomb-interactions-case (very low charge per bunch) could be recovered even for larger bunch charges using aberration free electron lenses. This gives us an upper limit to the brightness that can be obtained from ASP as shown by the green curve. The purple curve shows the brightness with the electrons/shot extrapolated in the range beyond 3 kW laser peak power. The gray area shows the region in which the brightness from ASPs could lie depending on the nature of the correlations developed in $x$ and $p_x$ due to the Coulomb interactions.

Fig. \ref{fgr:Graphs} (b) also shows the black cross (\textbf{$\times$}) symbol indicating the 4D-Brightness of the electron bunch obtained in the HiRES UED beamline from a flat cathode after using collimating apertures \cite{ji2019ultrafast}. It should be noted that this point indicates the brightness at the sample whereas the performance of the ASP is at the source.

To-date all electron sources using flat photocathodes have been operated with large $\mu$m-mm sized emission areas putting them in the regime where electron-electron Coulomb interactions can be modeled by assuming the electron bunch to be a continuous distribution of charge \cite{bazarov2009maximum,dowell2019topological}. In such cases, the accelerating electric field ($\mathscr{E}$) at the cathode limits the maximum charge density ($\sigma_{max}$) in pancake regime to $\sigma_{max} = \epsilon_0 \mathscr{E}$ \cite{bazarov2009maximum}. Here, $\epsilon_0$ is vacuum permittivity. With an electric field of 6 MV/m as obtained in the PEEM with the small rms spot size of $\approx$50 nm, we get the maximum charge that can be extracted from ASPs to be $\approx$2 electrons/shot within this space charge assumption.  

However, the space charge assumption breaks down for a few electrons and it is imperative to consider individual Coulomb interactions between each pair of electrons to a properly account for the Coulomb effects \cite{meier2023few}. As a result, extraction of more than two electrons may be possible even at these small electric fields. Furthermore extraction of higher charges per bunch could be possible with larger electric fields in the range of 20-100 MV$/$m that are typically found in RF guns \cite{faillace2014recent}. For instance, operating ASP in the HiRES UED beamline with the peak RF field of $\approx$20 MV/m, we can extract $\approx$6 electrons/shot within the space charge assumption discussed in the previous paragraph. Taking $\epsilon_{n,x} = 40$ pm-rad with 6 electrons/shot implies a maximum 4D-Brightness of $\approx$3700 electrons$/$(nm$^{2}$Sr) which is $\sim$40 times better than the current maximum 4D-Brightness achieved in the HiRES UED beamline at the sample plane \cite{li2022kiloelectron,siddiqui2023relativistic}.

It is also worth noting that the maximum laser fluence used in this experiment, after accounting for a factor of $\approx$300 in intensity enhancement is 5 mJ/cm$^2$.  This is 20 times lower than the damage threshold of gold which is typically in the range of 100-1000 mJ/cm$^2$ \cite{kruger2007femtosecond}. Even with 3 times higher laser power, due to the 5$^{th}$ order emission process one can expect 200 times more charge per shot and potentially an increased brightness. The actual increase in charge and the brightness will then be determined by the electron-electron interaction dynamics. 
 
%One of the potential impact of the nanostructured ASP, we compare their operational requirement with Schottky based photo-emitter used in Ref.\onlinecite{feist2017ultrafast}. The operation of Schottky based photoemitter requires a UV pulse energy of $\approx$10 nJ. Considering typical operational conversion efficiency of IR to UV of $\approx$7.5 \% \cite{pierce2023experimental}, 140 nJ of IR pulse energy is required for operation of Schottky based photo-emitter as compared to 0.4 nJ for ASP. 

Harnessing plasmonic focusing for the development of low emittance ultrafast ASP is the first step to develop next-generation state of the art ultra-bright electron sources for next-generation accelerator applications. ASP can be designed for an operating wavelength such that it can further reduce $\sigma_{p_{x}}$ and thereby $\epsilon_{n,x}$ of the emitted electron bunch.  In addition, ASP can be combined with high quantum efficiency semiconductor thin films \cite{saha2022physically,cultrera2016ultra} to further enhance the 4D-brightness of the emitted electron bunch. 
%In addition, ASP also carries the potential for various types of applications. Firstly, it can be used for stroboscopic MeV UED/M applications requiring electron bunches with high degree of coherence. Further, as the pico-meter scale emittance has been achieved right from the cathode without the need of any apertures or electrostatic lenses, it can lead to the development of next generation ultrafast and ultra-compact UED/M apparatus. 
Beyond the UED/M applications, ASPs can also be used as the electron source for dielectric laser accelerators for the development of next generation compact accelerators which require $<$1 nm-rad normalized transverse emittance \cite{england2014dielectric,simakov2017diamond}. Additionally, the ASP and other types of plasmonic apertures \cite{dai2021ultrafast} can be arranged in an array to generate transversely shaped electron beams for powering novel coherent x-ray light sources \cite{graves2020asu} as well as the next generation of beam driven dielectric wakefield accelerators \cite{conde2017research}.

In summary, we have fabricated and characterized a plasmonics based, geometrically flat, low emittance and ultrafast source of electron pulses suitable for UED/M experiments. We achieved a record low rms normalized transverse electron emittance of less than 40 pm-rad from a geometrically flat photocathode – nearly an order of magnitude lower than the best the emittance that has been achieved from a geometrically flat photocathodes \cite{li2012nanometer,jin2008super,maxson2017direct}. Such a plasmonics based low emittance electron source operating with IR light can impact a wide range of applications ranging from high repetition rate UED/M to next generation  particle accelerators.

\section{acknowledgement}
This work is supported by the NSF Center for Bright Beams under award PHY-1549132 and Department of Energy Office of Science under awards DE-SC0021092, and DE-SC0021213. C.M.P. acknowledges support from the US DOE SCGSR program. J.M was partially supported by U.S Department of Energy, Grant No. DE-SC0020144.

\section{Supplementary Information}
Supplementary Information Consists of: Mathematical equation for designing tilt compensated ASP, details about FDTD simulation, experimental details and details about real-space measurement of ASP using mercury (Hg) lamp
and pulsed femtosecond laser in conjunction.

\nocite{*}
\bibliography{Bibliography}% Produces the bibliography via BibTeX.

\begin{thebibliography}{63}%
\makeatletter
\providecommand \@ifxundefined [1]{%
 \@ifx{#1\undefined}
}%
\providecommand \@ifnum [1]{%
 \ifnum #1\expandafter \@firstoftwo
 \else \expandafter \@secondoftwo
 \fi
}%
\providecommand \@ifx [1]{%
 \ifx #1\expandafter \@firstoftwo
 \else \expandafter \@secondoftwo
 \fi
}%
\providecommand \natexlab [1]{#1}%
\providecommand \enquote  [1]{``#1''}%
\providecommand \bibnamefont  [1]{#1}%
\providecommand \bibfnamefont [1]{#1}%
\providecommand \citenamefont [1]{#1}%
\providecommand \href@noop [0]{\@secondoftwo}%
\providecommand \href [0]{\begingroup \@sanitize@url \@href}%
\providecommand \@href[1]{\@@startlink{#1}\@@href}%
\providecommand \@@href[1]{\endgroup#1\@@endlink}%
\providecommand \@sanitize@url [0]{\catcode `\\12\catcode `\$12\catcode `\&12\catcode `\#12\catcode `\^12\catcode `\_12\catcode `\%12\relax}%
\providecommand \@@startlink[1]{}%
\providecommand \@@endlink[0]{}%
\providecommand \url  [0]{\begingroup\@sanitize@url \@url }%
\providecommand \@url [1]{\endgroup\@href {#1}{\urlprefix }}%
\providecommand \urlprefix  [0]{URL }%
\providecommand \Eprint [0]{\href }%
\providecommand \doibase [0]{https://doi.org/}%
\providecommand \selectlanguage [0]{\@gobble}%
\providecommand \bibinfo  [0]{\@secondoftwo}%
\providecommand \bibfield  [0]{\@secondoftwo}%
\providecommand \translation [1]{[#1]}%
\providecommand \BibitemOpen [0]{}%
\providecommand \bibitemStop [0]{}%
\providecommand \bibitemNoStop [0]{.\EOS\space}%
\providecommand \EOS [0]{\spacefactor3000\relax}%
\providecommand \BibitemShut  [1]{\csname bibitem#1\endcsname}%
\let\auto@bib@innerbib\@empty
%</preamble>
\bibitem [{\citenamefont {Sood}\ \emph {et~al.}(2021)\citenamefont {Sood}, \citenamefont {Shen}, \citenamefont {Shi}, \citenamefont {Kumar}, \citenamefont {Park}, \citenamefont {Zajac}, \citenamefont {Sun}, \citenamefont {Chen}, \citenamefont {Ramanathan}, \citenamefont {Wang} \emph {et~al.}}]{sood2021universal}%
  \BibitemOpen
  \bibfield  {author} {\bibinfo {author} {\bibfnamefont {A.}~\bibnamefont {Sood}}, \bibinfo {author} {\bibfnamefont {X.}~\bibnamefont {Shen}}, \bibinfo {author} {\bibfnamefont {Y.}~\bibnamefont {Shi}}, \bibinfo {author} {\bibfnamefont {S.}~\bibnamefont {Kumar}}, \bibinfo {author} {\bibfnamefont {S.~J.}\ \bibnamefont {Park}}, \bibinfo {author} {\bibfnamefont {M.}~\bibnamefont {Zajac}}, \bibinfo {author} {\bibfnamefont {Y.}~\bibnamefont {Sun}}, \bibinfo {author} {\bibfnamefont {L.-Q.}\ \bibnamefont {Chen}}, \bibinfo {author} {\bibfnamefont {S.}~\bibnamefont {Ramanathan}}, \bibinfo {author} {\bibfnamefont {X.}~\bibnamefont {Wang}}, \emph {et~al.},\ }\bibfield  {title} {\bibinfo {title} {Universal phase dynamics in vo2 switches revealed by ultrafast operando diffraction},\ }\href@noop {} {\bibfield  {journal} {\bibinfo  {journal} {Science}\ }\textbf {\bibinfo {volume} {373}},\ \bibinfo {pages} {352} (\bibinfo {year} {2021})}\BibitemShut {NoStop}%
\bibitem [{\citenamefont {Siddiqui}\ \emph {et~al.}(2023)\citenamefont {Siddiqui}, \citenamefont {Durham}, \citenamefont {Cropp}, \citenamefont {Ji}, \citenamefont {Paiagua}, \citenamefont {Ophus}, \citenamefont {Andresen}, \citenamefont {Jin}, \citenamefont {Wu}, \citenamefont {Wang} \emph {et~al.}}]{siddiqui2023relativistic}%
  \BibitemOpen
  \bibfield  {author} {\bibinfo {author} {\bibfnamefont {K.}~\bibnamefont {Siddiqui}}, \bibinfo {author} {\bibfnamefont {D.}~\bibnamefont {Durham}}, \bibinfo {author} {\bibfnamefont {F.}~\bibnamefont {Cropp}}, \bibinfo {author} {\bibfnamefont {F.}~\bibnamefont {Ji}}, \bibinfo {author} {\bibfnamefont {S.}~\bibnamefont {Paiagua}}, \bibinfo {author} {\bibfnamefont {C.}~\bibnamefont {Ophus}}, \bibinfo {author} {\bibfnamefont {N.}~\bibnamefont {Andresen}}, \bibinfo {author} {\bibfnamefont {L.}~\bibnamefont {Jin}}, \bibinfo {author} {\bibfnamefont {J.}~\bibnamefont {Wu}}, \bibinfo {author} {\bibfnamefont {S.}~\bibnamefont {Wang}}, \emph {et~al.},\ }\bibfield  {title} {\bibinfo {title} {Relativistic ultrafast electron diffraction at high repetition rates},\ }\href@noop {} {\bibfield  {journal} {\bibinfo  {journal} {arXiv preprint arXiv:2306.04900}\ } (\bibinfo {year} {2023})}\BibitemShut {NoStop}%
\bibitem [{\citenamefont {Durham}\ \emph {et~al.}(2020)\citenamefont {Durham}, \citenamefont {Siddiqui}, \citenamefont {Ji}, \citenamefont {Navarro}, \citenamefont {Musumeci}, \citenamefont {Kaindl}, \citenamefont {Minor},\ and\ \citenamefont {Filippetto}}]{durham2020relativistic}%
  \BibitemOpen
  \bibfield  {author} {\bibinfo {author} {\bibfnamefont {D.}~\bibnamefont {Durham}}, \bibinfo {author} {\bibfnamefont {K.}~\bibnamefont {Siddiqui}}, \bibinfo {author} {\bibfnamefont {F.}~\bibnamefont {Ji}}, \bibinfo {author} {\bibfnamefont {J.~G.}\ \bibnamefont {Navarro}}, \bibinfo {author} {\bibfnamefont {P.}~\bibnamefont {Musumeci}}, \bibinfo {author} {\bibfnamefont {R.}~\bibnamefont {Kaindl}}, \bibinfo {author} {\bibfnamefont {A.}~\bibnamefont {Minor}},\ and\ \bibinfo {author} {\bibfnamefont {D.}~\bibnamefont {Filippetto}},\ }\bibfield  {title} {\bibinfo {title} {Relativistic ultrafast electron diffraction of nanomaterials},\ }\href@noop {} {\bibfield  {journal} {\bibinfo  {journal} {Microscopy and Microanalysis}\ }\textbf {\bibinfo {volume} {26}},\ \bibinfo {pages} {676} (\bibinfo {year} {2020})}\BibitemShut {NoStop}%
\bibitem [{\citenamefont {Gliserin}(2014)}]{gliserin2014towards}%
  \BibitemOpen
  \bibfield  {author} {\bibinfo {author} {\bibfnamefont {A.}~\bibnamefont {Gliserin}},\ }\emph {\bibinfo {title} {Towards attosecond 4D imaging of atomic-scale dynamics by single-electron diffraction}},\ \href@noop {} {Ph.D. thesis},\ \bibinfo  {school} {lmu} (\bibinfo {year} {2014})\BibitemShut {NoStop}%
\bibitem [{\citenamefont {Ji}\ \emph {et~al.}(2019)\citenamefont {Ji}, \citenamefont {Durham}, \citenamefont {Minor}, \citenamefont {Musumeci}, \citenamefont {Navarro},\ and\ \citenamefont {Filippetto}}]{ji2019ultrafast}%
  \BibitemOpen
  \bibfield  {author} {\bibinfo {author} {\bibfnamefont {F.}~\bibnamefont {Ji}}, \bibinfo {author} {\bibfnamefont {D.~B.}\ \bibnamefont {Durham}}, \bibinfo {author} {\bibfnamefont {A.~M.}\ \bibnamefont {Minor}}, \bibinfo {author} {\bibfnamefont {P.}~\bibnamefont {Musumeci}}, \bibinfo {author} {\bibfnamefont {J.~G.}\ \bibnamefont {Navarro}},\ and\ \bibinfo {author} {\bibfnamefont {D.}~\bibnamefont {Filippetto}},\ }\bibfield  {title} {\bibinfo {title} {Ultrafast relativistic electron nanoprobes},\ }\href@noop {} {\bibfield  {journal} {\bibinfo  {journal} {Communications Physics}\ }\textbf {\bibinfo {volume} {2}},\ \bibinfo {pages} {54} (\bibinfo {year} {2019})}\BibitemShut {NoStop}%
\bibitem [{\citenamefont {Ischenko}\ \emph {et~al.}(1983)\citenamefont {Ischenko}, \citenamefont {Golubkov}, \citenamefont {Spiridonov}, \citenamefont {Zgurskii}, \citenamefont {Akhmanov}, \citenamefont {Vabischevich},\ and\ \citenamefont {Bagratashvili}}]{ischenko1983stroboscopical}%
  \BibitemOpen
  \bibfield  {author} {\bibinfo {author} {\bibfnamefont {A.}~\bibnamefont {Ischenko}}, \bibinfo {author} {\bibfnamefont {V.}~\bibnamefont {Golubkov}}, \bibinfo {author} {\bibfnamefont {V.}~\bibnamefont {Spiridonov}}, \bibinfo {author} {\bibfnamefont {A.}~\bibnamefont {Zgurskii}}, \bibinfo {author} {\bibfnamefont {A.}~\bibnamefont {Akhmanov}}, \bibinfo {author} {\bibfnamefont {M.}~\bibnamefont {Vabischevich}},\ and\ \bibinfo {author} {\bibfnamefont {V.}~\bibnamefont {Bagratashvili}},\ }\bibfield  {title} {\bibinfo {title} {A stroboscopical gas-electron diffraction method for the investigation of short-lived molecular species},\ }\href@noop {} {\bibfield  {journal} {\bibinfo  {journal} {Applied Physics B}\ }\textbf {\bibinfo {volume} {32}},\ \bibinfo {pages} {161} (\bibinfo {year} {1983})}\BibitemShut {NoStop}%
\bibitem [{\citenamefont {Ischenko}\ \emph {et~al.}(1994)\citenamefont {Ischenko}, \citenamefont {Schafer}, \citenamefont {Luo},\ and\ \citenamefont {Ewbank}}]{ischenko1994structural}%
  \BibitemOpen
  \bibfield  {author} {\bibinfo {author} {\bibfnamefont {A.~A.}\ \bibnamefont {Ischenko}}, \bibinfo {author} {\bibfnamefont {L.}~\bibnamefont {Schafer}}, \bibinfo {author} {\bibfnamefont {J.~Y.}\ \bibnamefont {Luo}},\ and\ \bibinfo {author} {\bibfnamefont {J.~D.}\ \bibnamefont {Ewbank}},\ }\bibfield  {title} {\bibinfo {title} {Structural and vibrational kinetics by stroboscopic gas electron diffraction: The 193 nm photodissociation of cs2},\ }\href@noop {} {\bibfield  {journal} {\bibinfo  {journal} {The Journal of Physical Chemistry}\ }\textbf {\bibinfo {volume} {98}},\ \bibinfo {pages} {8673} (\bibinfo {year} {1994})}\BibitemShut {NoStop}%
\bibitem [{\citenamefont {Kirchner}(2013)}]{kirchner2013ultrashort}%
  \BibitemOpen
  \bibfield  {author} {\bibinfo {author} {\bibfnamefont {F.}~\bibnamefont {Kirchner}},\ }\emph {\bibinfo {title} {Ultrashort and coherent single-electron pulses for diffraction at ultimate resolutions}},\ \href@noop {} {Ph.D. thesis},\ \bibinfo  {school} {lmu} (\bibinfo {year} {2013})\BibitemShut {NoStop}%
\bibitem [{\citenamefont {Musumeci}\ \emph {et~al.}(2018)\citenamefont {Musumeci}, \citenamefont {Navarro}, \citenamefont {Rosenzweig}, \citenamefont {Cultrera}, \citenamefont {Bazarov}, \citenamefont {Maxson}, \citenamefont {Karkare},\ and\ \citenamefont {Padmore}}]{musumeci2018advances}%
  \BibitemOpen
  \bibfield  {author} {\bibinfo {author} {\bibfnamefont {P.}~\bibnamefont {Musumeci}}, \bibinfo {author} {\bibfnamefont {J.~G.}\ \bibnamefont {Navarro}}, \bibinfo {author} {\bibfnamefont {J.}~\bibnamefont {Rosenzweig}}, \bibinfo {author} {\bibfnamefont {L.}~\bibnamefont {Cultrera}}, \bibinfo {author} {\bibfnamefont {I.}~\bibnamefont {Bazarov}}, \bibinfo {author} {\bibfnamefont {J.}~\bibnamefont {Maxson}}, \bibinfo {author} {\bibfnamefont {S.}~\bibnamefont {Karkare}},\ and\ \bibinfo {author} {\bibfnamefont {H.}~\bibnamefont {Padmore}},\ }\bibfield  {title} {\bibinfo {title} {Advances in bright electron sources},\ }\href@noop {} {\bibfield  {journal} {\bibinfo  {journal} {Nuclear Instruments and Methods in Physics Research Section A: Accelerators, Spectrometers, Detectors and Associated Equipment}\ }\textbf {\bibinfo {volume} {907}},\ \bibinfo {pages} {209} (\bibinfo {year} {2018})}\BibitemShut {NoStop}%
\bibitem [{\citenamefont {Li}\ \emph {et~al.}(2022)\citenamefont {Li}, \citenamefont {Duncan}, \citenamefont {Andorf}, \citenamefont {Bartnik}, \citenamefont {Bianco}, \citenamefont {Cultrera}, \citenamefont {Galdi}, \citenamefont {Gordon}, \citenamefont {Kaemingk}, \citenamefont {Pennington} \emph {et~al.}}]{li2022kiloelectron}%
  \BibitemOpen
  \bibfield  {author} {\bibinfo {author} {\bibfnamefont {W.}~\bibnamefont {Li}}, \bibinfo {author} {\bibfnamefont {C.}~\bibnamefont {Duncan}}, \bibinfo {author} {\bibfnamefont {M.}~\bibnamefont {Andorf}}, \bibinfo {author} {\bibfnamefont {A.}~\bibnamefont {Bartnik}}, \bibinfo {author} {\bibfnamefont {E.}~\bibnamefont {Bianco}}, \bibinfo {author} {\bibfnamefont {L.}~\bibnamefont {Cultrera}}, \bibinfo {author} {\bibfnamefont {A.}~\bibnamefont {Galdi}}, \bibinfo {author} {\bibfnamefont {M.}~\bibnamefont {Gordon}}, \bibinfo {author} {\bibfnamefont {M.}~\bibnamefont {Kaemingk}}, \bibinfo {author} {\bibfnamefont {C.}~\bibnamefont {Pennington}}, \emph {et~al.},\ }\bibfield  {title} {\bibinfo {title} {A kiloelectron-volt ultrafast electron micro-diffraction apparatus using low emittance semiconductor photocathodes},\ }\href@noop {} {\bibfield  {journal} {\bibinfo  {journal} {Structural Dynamics}\ }\textbf {\bibinfo {volume} {9}} (\bibinfo {year} {2022})}\BibitemShut {NoStop}%
\bibitem [{\citenamefont {Filippetto}\ and\ \citenamefont {Qian}(2016)}]{filippetto2016design}%
  \BibitemOpen
  \bibfield  {author} {\bibinfo {author} {\bibfnamefont {D.}~\bibnamefont {Filippetto}}\ and\ \bibinfo {author} {\bibfnamefont {H.}~\bibnamefont {Qian}},\ }\bibfield  {title} {\bibinfo {title} {Design of a high-flux instrument for ultrafast electron diffraction and microscopy},\ }\href@noop {} {\bibfield  {journal} {\bibinfo  {journal} {Journal of Physics B: Atomic, Molecular and Optical Physics}\ }\textbf {\bibinfo {volume} {49}},\ \bibinfo {pages} {104003} (\bibinfo {year} {2016})}\BibitemShut {NoStop}%
\bibitem [{\citenamefont {Filippetto}\ \emph {et~al.}(2022)\citenamefont {Filippetto}, \citenamefont {Musumeci}, \citenamefont {Li}, \citenamefont {Siwick}, \citenamefont {Otto}, \citenamefont {Centurion},\ and\ \citenamefont {Nunes}}]{filippetto2022ultrafast}%
  \BibitemOpen
  \bibfield  {author} {\bibinfo {author} {\bibfnamefont {D.}~\bibnamefont {Filippetto}}, \bibinfo {author} {\bibfnamefont {P.}~\bibnamefont {Musumeci}}, \bibinfo {author} {\bibfnamefont {R.}~\bibnamefont {Li}}, \bibinfo {author} {\bibfnamefont {B.~J.}\ \bibnamefont {Siwick}}, \bibinfo {author} {\bibfnamefont {M.}~\bibnamefont {Otto}}, \bibinfo {author} {\bibfnamefont {M.}~\bibnamefont {Centurion}},\ and\ \bibinfo {author} {\bibfnamefont {J.}~\bibnamefont {Nunes}},\ }\bibfield  {title} {\bibinfo {title} {Ultrafast electron diffraction: Visualizing dynamic states of matter},\ }\href@noop {} {\bibfield  {journal} {\bibinfo  {journal} {Reviews of Modern Physics}\ }\textbf {\bibinfo {volume} {94}},\ \bibinfo {pages} {045004} (\bibinfo {year} {2022})}\BibitemShut {NoStop}%
\bibitem [{\citenamefont {Dowell}\ and\ \citenamefont {Schmerge}(2009)}]{dowell2009quantum}%
  \BibitemOpen
  \bibfield  {author} {\bibinfo {author} {\bibfnamefont {D.~H.}\ \bibnamefont {Dowell}}\ and\ \bibinfo {author} {\bibfnamefont {J.~F.}\ \bibnamefont {Schmerge}},\ }\bibfield  {title} {\bibinfo {title} {Quantum efficiency and thermal emittance of metal photocathodes},\ }\href@noop {} {\bibfield  {journal} {\bibinfo  {journal} {Physical Review Special Topics-Accelerators and Beams}\ }\textbf {\bibinfo {volume} {12}},\ \bibinfo {pages} {074201} (\bibinfo {year} {2009})}\BibitemShut {NoStop}%
\bibitem [{\citenamefont {Feist}\ \emph {et~al.}(2017)\citenamefont {Feist}, \citenamefont {Bach}, \citenamefont {da~Silva}, \citenamefont {Danz}, \citenamefont {M{\"o}ller}, \citenamefont {Priebe}, \citenamefont {Domr{\"o}se}, \citenamefont {Gatzmann}, \citenamefont {Rost}, \citenamefont {Schauss} \emph {et~al.}}]{feist2017ultrafast}%
  \BibitemOpen
  \bibfield  {author} {\bibinfo {author} {\bibfnamefont {A.}~\bibnamefont {Feist}}, \bibinfo {author} {\bibfnamefont {N.}~\bibnamefont {Bach}}, \bibinfo {author} {\bibfnamefont {N.~R.}\ \bibnamefont {da~Silva}}, \bibinfo {author} {\bibfnamefont {T.}~\bibnamefont {Danz}}, \bibinfo {author} {\bibfnamefont {M.}~\bibnamefont {M{\"o}ller}}, \bibinfo {author} {\bibfnamefont {K.~E.}\ \bibnamefont {Priebe}}, \bibinfo {author} {\bibfnamefont {T.}~\bibnamefont {Domr{\"o}se}}, \bibinfo {author} {\bibfnamefont {J.~G.}\ \bibnamefont {Gatzmann}}, \bibinfo {author} {\bibfnamefont {S.}~\bibnamefont {Rost}}, \bibinfo {author} {\bibfnamefont {J.}~\bibnamefont {Schauss}}, \emph {et~al.},\ }\bibfield  {title} {\bibinfo {title} {Ultrafast transmission electron microscopy using a laser-driven field emitter: Femtosecond resolution with a high coherence electron beam},\ }\href@noop {} {\bibfield  {journal} {\bibinfo  {journal} {Ultramicroscopy}\ }\textbf {\bibinfo {volume} {176}},\ \bibinfo {pages} {63} (\bibinfo {year}
  {2017})}\BibitemShut {NoStop}%
\bibitem [{\citenamefont {Chatelain}\ \emph {et~al.}(2012)\citenamefont {Chatelain}, \citenamefont {Morrison}, \citenamefont {Godbout},\ and\ \citenamefont {Siwick}}]{chatelain2012ultrafast}%
  \BibitemOpen
  \bibfield  {author} {\bibinfo {author} {\bibfnamefont {R.~P.}\ \bibnamefont {Chatelain}}, \bibinfo {author} {\bibfnamefont {V.~R.}\ \bibnamefont {Morrison}}, \bibinfo {author} {\bibfnamefont {C.}~\bibnamefont {Godbout}},\ and\ \bibinfo {author} {\bibfnamefont {B.~J.}\ \bibnamefont {Siwick}},\ }\bibfield  {title} {\bibinfo {title} {Ultrafast electron diffraction with radio-frequency compressed electron pulses},\ }\href@noop {} {\bibfield  {journal} {\bibinfo  {journal} {Applied Physics Letters}\ }\textbf {\bibinfo {volume} {101}} (\bibinfo {year} {2012})}\BibitemShut {NoStop}%
\bibitem [{\citenamefont {Weathersby}\ \emph {et~al.}(2015)\citenamefont {Weathersby}, \citenamefont {Brown}, \citenamefont {Centurion}, \citenamefont {Chase}, \citenamefont {Coffee}, \citenamefont {Corbett}, \citenamefont {Eichner}, \citenamefont {Frisch}, \citenamefont {Fry}, \citenamefont {G{\"u}hr} \emph {et~al.}}]{weathersby2015mega}%
  \BibitemOpen
  \bibfield  {author} {\bibinfo {author} {\bibfnamefont {S.}~\bibnamefont {Weathersby}}, \bibinfo {author} {\bibfnamefont {G.}~\bibnamefont {Brown}}, \bibinfo {author} {\bibfnamefont {M.}~\bibnamefont {Centurion}}, \bibinfo {author} {\bibfnamefont {T.}~\bibnamefont {Chase}}, \bibinfo {author} {\bibfnamefont {R.}~\bibnamefont {Coffee}}, \bibinfo {author} {\bibfnamefont {J.}~\bibnamefont {Corbett}}, \bibinfo {author} {\bibfnamefont {J.}~\bibnamefont {Eichner}}, \bibinfo {author} {\bibfnamefont {J.}~\bibnamefont {Frisch}}, \bibinfo {author} {\bibfnamefont {A.}~\bibnamefont {Fry}}, \bibinfo {author} {\bibfnamefont {M.}~\bibnamefont {G{\"u}hr}}, \emph {et~al.},\ }\bibfield  {title} {\bibinfo {title} {Mega-electron-volt ultrafast electron diffraction at slac national accelerator laboratory},\ }\href@noop {} {\bibfield  {journal} {\bibinfo  {journal} {Review of Scientific Instruments}\ }\textbf {\bibinfo {volume} {86}} (\bibinfo {year} {2015})}\BibitemShut {NoStop}%
\bibitem [{\citenamefont {Carter}\ and\ \citenamefont {Williams}(2016)}]{carter2016transmission}%
  \BibitemOpen
  \bibfield  {author} {\bibinfo {author} {\bibfnamefont {C.~B.}\ \bibnamefont {Carter}}\ and\ \bibinfo {author} {\bibfnamefont {D.~B.}\ \bibnamefont {Williams}},\ }\href@noop {} {\emph {\bibinfo {title} {Transmission electron microscopy: Diffraction, imaging, and spectrometry}}}\ (\bibinfo  {publisher} {Springer},\ \bibinfo {year} {2016})\BibitemShut {NoStop}%
\bibitem [{\citenamefont {Parzyck}\ \emph {et~al.}(2022)\citenamefont {Parzyck}, \citenamefont {Galdi}, \citenamefont {Nangoi}, \citenamefont {DeBenedetti}, \citenamefont {Balajka}, \citenamefont {Faeth}, \citenamefont {Paik}, \citenamefont {Hu}, \citenamefont {Arias}, \citenamefont {Hines}, \citenamefont {Schlom}, \citenamefont {Shen},\ and\ \citenamefont {Maxson}}]{PhysRevLett.128.114801}%
  \BibitemOpen
  \bibfield  {author} {\bibinfo {author} {\bibfnamefont {C.~T.}\ \bibnamefont {Parzyck}}, \bibinfo {author} {\bibfnamefont {A.}~\bibnamefont {Galdi}}, \bibinfo {author} {\bibfnamefont {J.~K.}\ \bibnamefont {Nangoi}}, \bibinfo {author} {\bibfnamefont {W.~J.~I.}\ \bibnamefont {DeBenedetti}}, \bibinfo {author} {\bibfnamefont {J.}~\bibnamefont {Balajka}}, \bibinfo {author} {\bibfnamefont {B.~D.}\ \bibnamefont {Faeth}}, \bibinfo {author} {\bibfnamefont {H.}~\bibnamefont {Paik}}, \bibinfo {author} {\bibfnamefont {C.}~\bibnamefont {Hu}}, \bibinfo {author} {\bibfnamefont {T.~A.}\ \bibnamefont {Arias}}, \bibinfo {author} {\bibfnamefont {M.~A.}\ \bibnamefont {Hines}}, \bibinfo {author} {\bibfnamefont {D.~G.}\ \bibnamefont {Schlom}}, \bibinfo {author} {\bibfnamefont {K.~M.}\ \bibnamefont {Shen}},\ and\ \bibinfo {author} {\bibfnamefont {J.~M.}\ \bibnamefont {Maxson}},\ }\bibfield  {title} {\bibinfo {title} {Single-crystal alkali antimonide photocathodes: High efficiency in the ultrathin limit},\ }\href
  {https://doi.org/10.1103/PhysRevLett.128.114801} {\bibfield  {journal} {\bibinfo  {journal} {Phys. Rev. Lett.}\ }\textbf {\bibinfo {volume} {128}},\ \bibinfo {pages} {114801} (\bibinfo {year} {2022})}\BibitemShut {NoStop}%
\bibitem [{\citenamefont {Galdi}\ \emph {et~al.}(2021)\citenamefont {Galdi}, \citenamefont {Balajka}, \citenamefont {DeBenedetti}, \citenamefont {Cultrera}, \citenamefont {Bazarov}, \citenamefont {Hines},\ and\ \citenamefont {Maxson}}]{galdi2021reduction}%
  \BibitemOpen
  \bibfield  {author} {\bibinfo {author} {\bibfnamefont {A.}~\bibnamefont {Galdi}}, \bibinfo {author} {\bibfnamefont {J.}~\bibnamefont {Balajka}}, \bibinfo {author} {\bibfnamefont {W.~J.}\ \bibnamefont {DeBenedetti}}, \bibinfo {author} {\bibfnamefont {L.}~\bibnamefont {Cultrera}}, \bibinfo {author} {\bibfnamefont {I.~V.}\ \bibnamefont {Bazarov}}, \bibinfo {author} {\bibfnamefont {M.~A.}\ \bibnamefont {Hines}},\ and\ \bibinfo {author} {\bibfnamefont {J.~M.}\ \bibnamefont {Maxson}},\ }\bibfield  {title} {\bibinfo {title} {Reduction of surface roughness emittance of cs3sb photocathodes grown via codeposition on single crystal substrates},\ }\href@noop {} {\bibfield  {journal} {\bibinfo  {journal} {Applied Physics Letters}\ }\textbf {\bibinfo {volume} {118}} (\bibinfo {year} {2021})}\BibitemShut {NoStop}%
\bibitem [{\citenamefont {Saha}\ \emph {et~al.}(2022)\citenamefont {Saha}, \citenamefont {Chubenko}, \citenamefont {Gevorkyan}, \citenamefont {Kachwala}, \citenamefont {Knill}, \citenamefont {Sarabia-Cardenas}, \citenamefont {Montgomery}, \citenamefont {Poddar}, \citenamefont {Paul}, \citenamefont {Hennig} \emph {et~al.}}]{saha2022physically}%
  \BibitemOpen
  \bibfield  {author} {\bibinfo {author} {\bibfnamefont {P.}~\bibnamefont {Saha}}, \bibinfo {author} {\bibfnamefont {O.}~\bibnamefont {Chubenko}}, \bibinfo {author} {\bibfnamefont {G.~S.}\ \bibnamefont {Gevorkyan}}, \bibinfo {author} {\bibfnamefont {A.}~\bibnamefont {Kachwala}}, \bibinfo {author} {\bibfnamefont {C.~J.}\ \bibnamefont {Knill}}, \bibinfo {author} {\bibfnamefont {C.}~\bibnamefont {Sarabia-Cardenas}}, \bibinfo {author} {\bibfnamefont {E.}~\bibnamefont {Montgomery}}, \bibinfo {author} {\bibfnamefont {S.}~\bibnamefont {Poddar}}, \bibinfo {author} {\bibfnamefont {J.~T.}\ \bibnamefont {Paul}}, \bibinfo {author} {\bibfnamefont {R.~G.}\ \bibnamefont {Hennig}}, \emph {et~al.},\ }\bibfield  {title} {\bibinfo {title} {Physically and chemically smooth cesium-antimonide photocathodes on single crystal strontium titanate substrates},\ }\href@noop {} {\bibfield  {journal} {\bibinfo  {journal} {Applied Physics Letters}\ }\textbf {\bibinfo {volume} {120}} (\bibinfo {year} {2022})}\BibitemShut {NoStop}%
\bibitem [{\citenamefont {Feng}\ \emph {et~al.}(2017)\citenamefont {Feng}, \citenamefont {Karkare}, \citenamefont {Nasiatka}, \citenamefont {Schubert}, \citenamefont {Smedley},\ and\ \citenamefont {Padmore}}]{feng2017near}%
  \BibitemOpen
  \bibfield  {author} {\bibinfo {author} {\bibfnamefont {J.}~\bibnamefont {Feng}}, \bibinfo {author} {\bibfnamefont {S.}~\bibnamefont {Karkare}}, \bibinfo {author} {\bibfnamefont {J.}~\bibnamefont {Nasiatka}}, \bibinfo {author} {\bibfnamefont {S.}~\bibnamefont {Schubert}}, \bibinfo {author} {\bibfnamefont {J.}~\bibnamefont {Smedley}},\ and\ \bibinfo {author} {\bibfnamefont {H.}~\bibnamefont {Padmore}},\ }\bibfield  {title} {\bibinfo {title} {Near atomically smooth alkali antimonide photocathode thin films},\ }\href@noop {} {\bibfield  {journal} {\bibinfo  {journal} {Journal of Applied Physics}\ }\textbf {\bibinfo {volume} {121}} (\bibinfo {year} {2017})}\BibitemShut {NoStop}%
\bibitem [{\citenamefont {Cultrera}\ \emph {et~al.}(2016)\citenamefont {Cultrera}, \citenamefont {Gulliford}, \citenamefont {Bartnik}, \citenamefont {Lee},\ and\ \citenamefont {Bazarov}}]{cultrera2016ultra}%
  \BibitemOpen
  \bibfield  {author} {\bibinfo {author} {\bibfnamefont {L.}~\bibnamefont {Cultrera}}, \bibinfo {author} {\bibfnamefont {C.}~\bibnamefont {Gulliford}}, \bibinfo {author} {\bibfnamefont {A.}~\bibnamefont {Bartnik}}, \bibinfo {author} {\bibfnamefont {H.}~\bibnamefont {Lee}},\ and\ \bibinfo {author} {\bibfnamefont {I.}~\bibnamefont {Bazarov}},\ }\bibfield  {title} {\bibinfo {title} {Ultra low emittance electron beams from multi-alkali antimonide photocathode operated with infrared light},\ }\href@noop {} {\bibfield  {journal} {\bibinfo  {journal} {Applied Physics Letters}\ }\textbf {\bibinfo {volume} {108}} (\bibinfo {year} {2016})}\BibitemShut {NoStop}%
\bibitem [{\citenamefont {Kachwala}\ \emph {et~al.}(2023)\citenamefont {Kachwala}, \citenamefont {Saha}, \citenamefont {Bhattacharyya}, \citenamefont {Montgomery}, \citenamefont {Chubenko},\ and\ \citenamefont {Karkare}}]{kachwala2023demonstration}%
  \BibitemOpen
  \bibfield  {author} {\bibinfo {author} {\bibfnamefont {A.}~\bibnamefont {Kachwala}}, \bibinfo {author} {\bibfnamefont {P.}~\bibnamefont {Saha}}, \bibinfo {author} {\bibfnamefont {P.}~\bibnamefont {Bhattacharyya}}, \bibinfo {author} {\bibfnamefont {E.}~\bibnamefont {Montgomery}}, \bibinfo {author} {\bibfnamefont {O.}~\bibnamefont {Chubenko}},\ and\ \bibinfo {author} {\bibfnamefont {S.}~\bibnamefont {Karkare}},\ }\bibfield  {title} {\bibinfo {title} {Demonstration of thermal limit mean transverse energy from cesium antimonide photocathodes},\ }\href@noop {} {\bibfield  {journal} {\bibinfo  {journal} {Applied Physics Letters}\ }\textbf {\bibinfo {volume} {123}} (\bibinfo {year} {2023})}\BibitemShut {NoStop}%
\bibitem [{\citenamefont {Saha}\ \emph {et~al.}(2021)\citenamefont {Saha}, \citenamefont {Chubenko}, \citenamefont {Gevorkyan}, \citenamefont {Kachwala}, \citenamefont {Karkare}, \citenamefont {Knill}, \citenamefont {Montgomery}, \citenamefont {Padmore}, \citenamefont {Poddar} \emph {et~al.}}]{saha2021optical}%
  \BibitemOpen
  \bibfield  {author} {\bibinfo {author} {\bibfnamefont {P.}~\bibnamefont {Saha}}, \bibinfo {author} {\bibfnamefont {O.}~\bibnamefont {Chubenko}}, \bibinfo {author} {\bibfnamefont {G.}~\bibnamefont {Gevorkyan}}, \bibinfo {author} {\bibfnamefont {A.}~\bibnamefont {Kachwala}}, \bibinfo {author} {\bibfnamefont {S.}~\bibnamefont {Karkare}}, \bibinfo {author} {\bibfnamefont {C.}~\bibnamefont {Knill}}, \bibinfo {author} {\bibfnamefont {E.}~\bibnamefont {Montgomery}}, \bibinfo {author} {\bibfnamefont {H.}~\bibnamefont {Padmore}}, \bibinfo {author} {\bibfnamefont {S.}~\bibnamefont {Poddar}}, \emph {et~al.},\ }\bibfield  {title} {\bibinfo {title} {Optical and surface characterization of alkali-antimonide photocathodes},\ }in\ \href@noop {} {\emph {\bibinfo {booktitle} {12th International Particle Accelerator Conference (IPAC'21), Campinas, SP, Brazil, 24-28 May 2021}}}\ (\bibinfo {organization} {JACOW Publishing, Geneva, Switzerland},\ \bibinfo {year} {2021})\ pp.\ \bibinfo {pages} {4037--4040}\BibitemShut {NoStop}%
\bibitem [{\citenamefont {Knill}\ \emph {et~al.}(2023{\natexlab{a}})\citenamefont {Knill}, \citenamefont {Yamaguchi}, \citenamefont {Kawahara}, \citenamefont {Wang}, \citenamefont {Batista}, \citenamefont {Yang}, \citenamefont {Ago}, \citenamefont {Moody},\ and\ \citenamefont {Karkare}}]{PhysRevApplied.19.014015}%
  \BibitemOpen
  \bibfield  {author} {\bibinfo {author} {\bibfnamefont {C.~J.}\ \bibnamefont {Knill}}, \bibinfo {author} {\bibfnamefont {H.}~\bibnamefont {Yamaguchi}}, \bibinfo {author} {\bibfnamefont {K.}~\bibnamefont {Kawahara}}, \bibinfo {author} {\bibfnamefont {G.}~\bibnamefont {Wang}}, \bibinfo {author} {\bibfnamefont {E.}~\bibnamefont {Batista}}, \bibinfo {author} {\bibfnamefont {P.}~\bibnamefont {Yang}}, \bibinfo {author} {\bibfnamefont {H.}~\bibnamefont {Ago}}, \bibinfo {author} {\bibfnamefont {N.}~\bibnamefont {Moody}},\ and\ \bibinfo {author} {\bibfnamefont {S.}~\bibnamefont {Karkare}},\ }\bibfield  {title} {\bibinfo {title} {Near-threshold photoemission from graphene-coated $\mathrm{Cu}$(110)},\ }\href {https://doi.org/10.1103/PhysRevApplied.19.014015} {\bibfield  {journal} {\bibinfo  {journal} {Phys. Rev. Appl.}\ }\textbf {\bibinfo {volume} {19}},\ \bibinfo {pages} {014015} (\bibinfo {year} {2023}{\natexlab{a}})}\BibitemShut {NoStop}%
\bibitem [{\citenamefont {Knill}(2023)}]{knill2023practical}%
  \BibitemOpen
  \bibfield  {author} {\bibinfo {author} {\bibfnamefont {C.~J.}\ \bibnamefont {Knill}},\ }\href@noop {} {\emph {\bibinfo {title} {Practical Limitations of Low Mean Transverse Energy Metallic Photocathodes}}},\ \bibinfo {type} {Tech. Rep.}\ (\bibinfo  {institution} {Arizona State University},\ \bibinfo {year} {2023})\BibitemShut {NoStop}%
\bibitem [{\citenamefont {Soomary}\ \emph {et~al.}(2021)\citenamefont {Soomary}, \citenamefont {Juarez-Lopez}, \citenamefont {Welsch}, \citenamefont {Jones},\ and\ \citenamefont {Noakes}}]{soomary2021performance}%
  \BibitemOpen
  \bibfield  {author} {\bibinfo {author} {\bibfnamefont {L.}~\bibnamefont {Soomary}}, \bibinfo {author} {\bibfnamefont {D.}~\bibnamefont {Juarez-Lopez}}, \bibinfo {author} {\bibfnamefont {C.}~\bibnamefont {Welsch}}, \bibinfo {author} {\bibfnamefont {L.}~\bibnamefont {Jones}},\ and\ \bibinfo {author} {\bibfnamefont {T.}~\bibnamefont {Noakes}},\ }\bibfield  {title} {\bibinfo {title} {Performance characterisation of a cu (100) single--crystal photocathode},\ }in\ \href@noop {} {\emph {\bibinfo {booktitle} {Proceedings of IPAC 2021 conference, Campinas, Brazil}}}\ (\bibinfo {year} {2021})\ pp.\ \bibinfo {pages} {2860--2862}\BibitemShut {NoStop}%
\bibitem [{\citenamefont {Knill}\ \emph {et~al.}(2023{\natexlab{b}})\citenamefont {Knill}, \citenamefont {Douyon}, \citenamefont {Kawahara}, \citenamefont {Yamaguchi}, \citenamefont {Wang}, \citenamefont {Ago}, \citenamefont {Moody},\ and\ \citenamefont {Karkare}}]{PhysRevAccelBeams.26.093401}%
  \BibitemOpen
  \bibfield  {author} {\bibinfo {author} {\bibfnamefont {C.~J.}\ \bibnamefont {Knill}}, \bibinfo {author} {\bibfnamefont {S.}~\bibnamefont {Douyon}}, \bibinfo {author} {\bibfnamefont {K.}~\bibnamefont {Kawahara}}, \bibinfo {author} {\bibfnamefont {H.}~\bibnamefont {Yamaguchi}}, \bibinfo {author} {\bibfnamefont {G.}~\bibnamefont {Wang}}, \bibinfo {author} {\bibfnamefont {H.}~\bibnamefont {Ago}}, \bibinfo {author} {\bibfnamefont {N.}~\bibnamefont {Moody}},\ and\ \bibinfo {author} {\bibfnamefont {S.}~\bibnamefont {Karkare}},\ }\bibfield  {title} {\bibinfo {title} {Effects of nonlinear photoemission on mean transverse energy from metal photocathodes},\ }\href {https://doi.org/10.1103/PhysRevAccelBeams.26.093401} {\bibfield  {journal} {\bibinfo  {journal} {Phys. Rev. Accel. Beams}\ }\textbf {\bibinfo {volume} {26}},\ \bibinfo {pages} {093401} (\bibinfo {year} {2023}{\natexlab{b}})}\BibitemShut {NoStop}%
\bibitem [{\citenamefont {Karkare}\ \emph {et~al.}(2014)\citenamefont {Karkare}, \citenamefont {Boulet}, \citenamefont {Cultrera}, \citenamefont {Dunham}, \citenamefont {Liu}, \citenamefont {Schaff},\ and\ \citenamefont {Bazarov}}]{karkare2014ultrabright}%
  \BibitemOpen
  \bibfield  {author} {\bibinfo {author} {\bibfnamefont {S.}~\bibnamefont {Karkare}}, \bibinfo {author} {\bibfnamefont {L.}~\bibnamefont {Boulet}}, \bibinfo {author} {\bibfnamefont {L.}~\bibnamefont {Cultrera}}, \bibinfo {author} {\bibfnamefont {B.}~\bibnamefont {Dunham}}, \bibinfo {author} {\bibfnamefont {X.}~\bibnamefont {Liu}}, \bibinfo {author} {\bibfnamefont {W.}~\bibnamefont {Schaff}},\ and\ \bibinfo {author} {\bibfnamefont {I.}~\bibnamefont {Bazarov}},\ }\bibfield  {title} {\bibinfo {title} {Ultrabright and ultrafast iii--v semiconductor photocathodes},\ }\href@noop {} {\bibfield  {journal} {\bibinfo  {journal} {Physical review letters}\ }\textbf {\bibinfo {volume} {112}},\ \bibinfo {pages} {097601} (\bibinfo {year} {2014})}\BibitemShut {NoStop}%
\bibitem [{\citenamefont {Karkare}\ \emph {et~al.}(2020)\citenamefont {Karkare}, \citenamefont {Adhikari}, \citenamefont {Schroeder}, \citenamefont {Nangoi}, \citenamefont {Arias}, \citenamefont {Maxson},\ and\ \citenamefont {Padmore}}]{karkare2020ultracold}%
  \BibitemOpen
  \bibfield  {author} {\bibinfo {author} {\bibfnamefont {S.}~\bibnamefont {Karkare}}, \bibinfo {author} {\bibfnamefont {G.}~\bibnamefont {Adhikari}}, \bibinfo {author} {\bibfnamefont {W.~A.}\ \bibnamefont {Schroeder}}, \bibinfo {author} {\bibfnamefont {J.~K.}\ \bibnamefont {Nangoi}}, \bibinfo {author} {\bibfnamefont {T.}~\bibnamefont {Arias}}, \bibinfo {author} {\bibfnamefont {J.}~\bibnamefont {Maxson}},\ and\ \bibinfo {author} {\bibfnamefont {H.}~\bibnamefont {Padmore}},\ }\bibfield  {title} {\bibinfo {title} {Ultracold electrons via near-threshold photoemission from single-crystal cu (100)},\ }\href@noop {} {\bibfield  {journal} {\bibinfo  {journal} {Physical review letters}\ }\textbf {\bibinfo {volume} {125}},\ \bibinfo {pages} {054801} (\bibinfo {year} {2020})}\BibitemShut {NoStop}%
\bibitem [{\citenamefont {Silfies}\ \emph {et~al.}(2019)\citenamefont {Silfies}, \citenamefont {Schwartz},\ and\ \citenamefont {Davidson}}]{silfies2019diffraction}%
  \BibitemOpen
  \bibfield  {author} {\bibinfo {author} {\bibfnamefont {J.~S.}\ \bibnamefont {Silfies}}, \bibinfo {author} {\bibfnamefont {S.~A.}\ \bibnamefont {Schwartz}},\ and\ \bibinfo {author} {\bibfnamefont {M.~W.}\ \bibnamefont {Davidson}},\ }\bibfield  {title} {\bibinfo {title} {The diffraction barrier in optical microscopy},\ }\href@noop {} {\bibfield  {journal} {\bibinfo  {journal} {Nikon Inc.,[Online]. Available: https://www. microscopyu. com/articles/superresolution/diffractionbarrier. html.[Accessed 28. 10. 2015]}\ } (\bibinfo {year} {2019})}\BibitemShut {NoStop}%
\bibitem [{\citenamefont {Li}\ \emph {et~al.}(2012)\citenamefont {Li}, \citenamefont {Roberts}, \citenamefont {Scoby}, \citenamefont {To},\ and\ \citenamefont {Musumeci}}]{li2012nanometer}%
  \BibitemOpen
  \bibfield  {author} {\bibinfo {author} {\bibfnamefont {R.}~\bibnamefont {Li}}, \bibinfo {author} {\bibfnamefont {K.}~\bibnamefont {Roberts}}, \bibinfo {author} {\bibfnamefont {C.}~\bibnamefont {Scoby}}, \bibinfo {author} {\bibfnamefont {H.}~\bibnamefont {To}},\ and\ \bibinfo {author} {\bibfnamefont {P.}~\bibnamefont {Musumeci}},\ }\bibfield  {title} {\bibinfo {title} {Nanometer emittance ultralow charge beams from rf photoinjectors},\ }\href@noop {} {\bibfield  {journal} {\bibinfo  {journal} {Physical Review Special Topics-Accelerators and Beams}\ }\textbf {\bibinfo {volume} {15}},\ \bibinfo {pages} {090702} (\bibinfo {year} {2012})}\BibitemShut {NoStop}%
\bibitem [{\citenamefont {Wen-Xi}\ \emph {et~al.}(2009)\citenamefont {Wen-Xi}, \citenamefont {Peng-Fei}, \citenamefont {Xuan}, \citenamefont {Shou-Hua}, \citenamefont {Zhong-Chao}, \citenamefont {Rick}, \citenamefont {Jian-Ming}, \citenamefont {Zheng-Ming},\ and\ \citenamefont {Jie}}]{wen2009ultrafast}%
  \BibitemOpen
  \bibfield  {author} {\bibinfo {author} {\bibfnamefont {L.}~\bibnamefont {Wen-Xi}}, \bibinfo {author} {\bibfnamefont {Z.}~\bibnamefont {Peng-Fei}}, \bibinfo {author} {\bibfnamefont {W.}~\bibnamefont {Xuan}}, \bibinfo {author} {\bibfnamefont {N.}~\bibnamefont {Shou-Hua}}, \bibinfo {author} {\bibfnamefont {Z.}~\bibnamefont {Zhong-Chao}}, \bibinfo {author} {\bibfnamefont {C.}~\bibnamefont {Rick}}, \bibinfo {author} {\bibfnamefont {C.}~\bibnamefont {Jian-Ming}}, \bibinfo {author} {\bibfnamefont {S.}~\bibnamefont {Zheng-Ming}},\ and\ \bibinfo {author} {\bibfnamefont {Z.}~\bibnamefont {Jie}},\ }\bibfield  {title} {\bibinfo {title} {Ultrafast electron diffraction with spatiotemporal resolution of atomic motion},\ }\href@noop {} {\bibfield  {journal} {\bibinfo  {journal} {Chinese Physics Letters}\ }\textbf {\bibinfo {volume} {26}},\ \bibinfo {pages} {020701} (\bibinfo {year} {2009})}\BibitemShut {NoStop}%
\bibitem [{\citenamefont {Gramotnev}\ and\ \citenamefont {Bozhevolnyi}(2010)}]{gramotnev2010plasmonics}%
  \BibitemOpen
  \bibfield  {author} {\bibinfo {author} {\bibfnamefont {D.~K.}\ \bibnamefont {Gramotnev}}\ and\ \bibinfo {author} {\bibfnamefont {S.~I.}\ \bibnamefont {Bozhevolnyi}},\ }\bibfield  {title} {\bibinfo {title} {Plasmonics beyond the diffraction limit},\ }\href@noop {} {\bibfield  {journal} {\bibinfo  {journal} {Nature photonics}\ }\textbf {\bibinfo {volume} {4}},\ \bibinfo {pages} {83} (\bibinfo {year} {2010})}\BibitemShut {NoStop}%
\bibitem [{\citenamefont {Pierce}\ \emph {et~al.}(2022)\citenamefont {Pierce}, \citenamefont {Daniel}, \citenamefont {Kachwala}, \citenamefont {Karkare}, \citenamefont {Bazarov}, \citenamefont {Maxson}, \citenamefont {Minor} \emph {et~al.}}]{pierce2022towards}%
  \BibitemOpen
  \bibfield  {author} {\bibinfo {author} {\bibfnamefont {C.~M.}\ \bibnamefont {Pierce}}, \bibinfo {author} {\bibfnamefont {B.}~\bibnamefont {Daniel}}, \bibinfo {author} {\bibfnamefont {A.}~\bibnamefont {Kachwala}}, \bibinfo {author} {\bibfnamefont {S.}~\bibnamefont {Karkare}}, \bibinfo {author} {\bibfnamefont {I.}~\bibnamefont {Bazarov}}, \bibinfo {author} {\bibfnamefont {J.}~\bibnamefont {Maxson}}, \bibinfo {author} {\bibfnamefont {A.~M.}\ \bibnamefont {Minor}}, \emph {et~al.},\ }\bibfield  {title} {\bibinfo {title} {Towards high brightness from plasmon-enhanced photoemitters},\ }in\ \href@noop {} {\emph {\bibinfo {booktitle} {Proceedings of the 5th North American Particle Accelerator Conference}}}\ (\bibinfo {year} {2022})\BibitemShut {NoStop}%
\bibitem [{\citenamefont {Durham}\ \emph {et~al.}(2019)\citenamefont {Durham}, \citenamefont {Riminucci}, \citenamefont {Ciabattini}, \citenamefont {Mostacci}, \citenamefont {Minor}, \citenamefont {Cabrini},\ and\ \citenamefont {Filippetto}}]{durham2019plasmonic}%
  \BibitemOpen
  \bibfield  {author} {\bibinfo {author} {\bibfnamefont {D.~B.}\ \bibnamefont {Durham}}, \bibinfo {author} {\bibfnamefont {F.}~\bibnamefont {Riminucci}}, \bibinfo {author} {\bibfnamefont {F.}~\bibnamefont {Ciabattini}}, \bibinfo {author} {\bibfnamefont {A.}~\bibnamefont {Mostacci}}, \bibinfo {author} {\bibfnamefont {A.~M.}\ \bibnamefont {Minor}}, \bibinfo {author} {\bibfnamefont {S.}~\bibnamefont {Cabrini}},\ and\ \bibinfo {author} {\bibfnamefont {D.}~\bibnamefont {Filippetto}},\ }\bibfield  {title} {\bibinfo {title} {Plasmonic lenses for tunable ultrafast electron emitters at the nanoscale},\ }\href@noop {} {\bibfield  {journal} {\bibinfo  {journal} {Physical Review Applied}\ }\textbf {\bibinfo {volume} {12}},\ \bibinfo {pages} {054057} (\bibinfo {year} {2019})}\BibitemShut {NoStop}%
\bibitem [{\citenamefont {et~al.}(2023)}]{kachwala:ipac2023-tupl188}%
  \BibitemOpen
  \bibfield  {author} {\bibinfo {author} {\bibfnamefont {A.~K.}\ \bibnamefont {et~al.}},\ }\bibfield  {title} {{\selectlanguage {English}\bibinfo {title} {Study of nano-structured electron sources using photoemission electron microscope}},\ }in\ \href {https://doi.org/10.18429/JACoW-IPAC2023-TUPL188} {{\selectlanguage {English}\emph {\bibinfo {booktitle} {Proc. 14th International Particle Accelerator Conference}}}},\ \bibinfo {series and number} {\bibinfo {series} {IPAC'23 - 14th International Particle Accelerator Conference}\ No.~\bibinfo {number} {14}}\ (\bibinfo  {publisher} {JACoW Publishing, Geneva, Switzerland},\ \bibinfo {year} {2023})\ pp.\ \bibinfo {pages} {2174--2177}\BibitemShut {NoStop}%
\bibitem [{\citenamefont {Guo}\ \emph {et~al.}(2017)\citenamefont {Guo}, \citenamefont {Li}, \citenamefont {Zhang}, \citenamefont {Guo}, \citenamefont {Shen}, \citenamefont {Zhou},\ and\ \citenamefont {Zhou}}]{guo2017review}%
  \BibitemOpen
  \bibfield  {author} {\bibinfo {author} {\bibfnamefont {Z.}~\bibnamefont {Guo}}, \bibinfo {author} {\bibfnamefont {Z.}~\bibnamefont {Li}}, \bibinfo {author} {\bibfnamefont {J.}~\bibnamefont {Zhang}}, \bibinfo {author} {\bibfnamefont {K.}~\bibnamefont {Guo}}, \bibinfo {author} {\bibfnamefont {F.}~\bibnamefont {Shen}}, \bibinfo {author} {\bibfnamefont {Q.}~\bibnamefont {Zhou}},\ and\ \bibinfo {author} {\bibfnamefont {H.}~\bibnamefont {Zhou}},\ }\bibfield  {title} {\bibinfo {title} {Review of the functions of archimedes’ spiral metallic nanostructures},\ }\href@noop {} {\bibfield  {journal} {\bibinfo  {journal} {Nanomaterials}\ }\textbf {\bibinfo {volume} {7}},\ \bibinfo {pages} {405} (\bibinfo {year} {2017})}\BibitemShut {NoStop}%
\bibitem [{\citenamefont {Tan}\ \emph {et~al.}(2017)\citenamefont {Tan}, \citenamefont {Argondizzo}, \citenamefont {Ren}, \citenamefont {Liu}, \citenamefont {Zhao},\ and\ \citenamefont {Petek}}]{tan2017plasmonic}%
  \BibitemOpen
  \bibfield  {author} {\bibinfo {author} {\bibfnamefont {S.}~\bibnamefont {Tan}}, \bibinfo {author} {\bibfnamefont {A.}~\bibnamefont {Argondizzo}}, \bibinfo {author} {\bibfnamefont {J.}~\bibnamefont {Ren}}, \bibinfo {author} {\bibfnamefont {L.}~\bibnamefont {Liu}}, \bibinfo {author} {\bibfnamefont {J.}~\bibnamefont {Zhao}},\ and\ \bibinfo {author} {\bibfnamefont {H.}~\bibnamefont {Petek}},\ }\bibfield  {title} {\bibinfo {title} {Plasmonic coupling at a metal/semiconductor interface},\ }\href@noop {} {\bibfield  {journal} {\bibinfo  {journal} {Nature Photonics}\ }\textbf {\bibinfo {volume} {11}},\ \bibinfo {pages} {806} (\bibinfo {year} {2017})}\BibitemShut {NoStop}%
\bibitem [{\citenamefont {Chen}\ \emph {et~al.}(2010)\citenamefont {Chen}, \citenamefont {Abeysinghe}, \citenamefont {Nelson},\ and\ \citenamefont {Zhan}}]{chen2010experimental}%
  \BibitemOpen
  \bibfield  {author} {\bibinfo {author} {\bibfnamefont {W.}~\bibnamefont {Chen}}, \bibinfo {author} {\bibfnamefont {D.~C.}\ \bibnamefont {Abeysinghe}}, \bibinfo {author} {\bibfnamefont {R.~L.}\ \bibnamefont {Nelson}},\ and\ \bibinfo {author} {\bibfnamefont {Q.}~\bibnamefont {Zhan}},\ }\bibfield  {title} {\bibinfo {title} {Experimental confirmation of miniature spiral plasmonic lens as a circular polarization analyzer},\ }\href@noop {} {\bibfield  {journal} {\bibinfo  {journal} {Nano letters}\ }\textbf {\bibinfo {volume} {10}},\ \bibinfo {pages} {2075} (\bibinfo {year} {2010})}\BibitemShut {NoStop}%
\bibitem [{\citenamefont {Jin}\ \emph {et~al.}(2008)\citenamefont {Jin}, \citenamefont {Yamamoto}, \citenamefont {Nakagawa}, \citenamefont {Mano}, \citenamefont {Kato}, \citenamefont {Tanioku}, \citenamefont {Ujihara}, \citenamefont {Takeda}, \citenamefont {Okumi}, \citenamefont {Yamamoto} \emph {et~al.}}]{jin2008super}%
  \BibitemOpen
  \bibfield  {author} {\bibinfo {author} {\bibfnamefont {X.}~\bibnamefont {Jin}}, \bibinfo {author} {\bibfnamefont {N.}~\bibnamefont {Yamamoto}}, \bibinfo {author} {\bibfnamefont {Y.}~\bibnamefont {Nakagawa}}, \bibinfo {author} {\bibfnamefont {A.}~\bibnamefont {Mano}}, \bibinfo {author} {\bibfnamefont {T.}~\bibnamefont {Kato}}, \bibinfo {author} {\bibfnamefont {M.}~\bibnamefont {Tanioku}}, \bibinfo {author} {\bibfnamefont {T.}~\bibnamefont {Ujihara}}, \bibinfo {author} {\bibfnamefont {Y.}~\bibnamefont {Takeda}}, \bibinfo {author} {\bibfnamefont {S.}~\bibnamefont {Okumi}}, \bibinfo {author} {\bibfnamefont {M.}~\bibnamefont {Yamamoto}}, \emph {et~al.},\ }\bibfield  {title} {\bibinfo {title} {Super-high brightness and high-spin-polarization photocathode},\ }\href@noop {} {\bibfield  {journal} {\bibinfo  {journal} {Applied physics express}\ }\textbf {\bibinfo {volume} {1}},\ \bibinfo {pages} {045002} (\bibinfo {year} {2008})}\BibitemShut {NoStop}%
\bibitem [{\citenamefont {Maxson}\ \emph {et~al.}(2017)\citenamefont {Maxson}, \citenamefont {Cesar}, \citenamefont {Calmasini}, \citenamefont {Ody}, \citenamefont {Musumeci},\ and\ \citenamefont {Alesini}}]{maxson2017direct}%
  \BibitemOpen
  \bibfield  {author} {\bibinfo {author} {\bibfnamefont {J.}~\bibnamefont {Maxson}}, \bibinfo {author} {\bibfnamefont {D.}~\bibnamefont {Cesar}}, \bibinfo {author} {\bibfnamefont {G.}~\bibnamefont {Calmasini}}, \bibinfo {author} {\bibfnamefont {A.}~\bibnamefont {Ody}}, \bibinfo {author} {\bibfnamefont {P.}~\bibnamefont {Musumeci}},\ and\ \bibinfo {author} {\bibfnamefont {D.}~\bibnamefont {Alesini}},\ }\bibfield  {title} {\bibinfo {title} {Direct measurement of sub-10 fs relativistic electron beams with ultralow emittance},\ }\href@noop {} {\bibfield  {journal} {\bibinfo  {journal} {Physical review letters}\ }\textbf {\bibinfo {volume} {118}},\ \bibinfo {pages} {154802} (\bibinfo {year} {2017})}\BibitemShut {NoStop}%
\bibitem [{\citenamefont {Frank}\ \emph {et~al.}(2017)\citenamefont {Frank}, \citenamefont {Kahl}, \citenamefont {Podbiel}, \citenamefont {Spektor}, \citenamefont {Orenstein}, \citenamefont {Fu}, \citenamefont {Weiss}, \citenamefont {Horn-von Hoegen}, \citenamefont {Davis}, \citenamefont {Meyer~zu Heringdorf} \emph {et~al.}}]{frank2017short}%
  \BibitemOpen
  \bibfield  {author} {\bibinfo {author} {\bibfnamefont {B.}~\bibnamefont {Frank}}, \bibinfo {author} {\bibfnamefont {P.}~\bibnamefont {Kahl}}, \bibinfo {author} {\bibfnamefont {D.}~\bibnamefont {Podbiel}}, \bibinfo {author} {\bibfnamefont {G.}~\bibnamefont {Spektor}}, \bibinfo {author} {\bibfnamefont {M.}~\bibnamefont {Orenstein}}, \bibinfo {author} {\bibfnamefont {L.}~\bibnamefont {Fu}}, \bibinfo {author} {\bibfnamefont {T.}~\bibnamefont {Weiss}}, \bibinfo {author} {\bibfnamefont {M.}~\bibnamefont {Horn-von Hoegen}}, \bibinfo {author} {\bibfnamefont {T.~J.}\ \bibnamefont {Davis}}, \bibinfo {author} {\bibfnamefont {F.-J.}\ \bibnamefont {Meyer~zu Heringdorf}}, \emph {et~al.},\ }\bibfield  {title} {\bibinfo {title} {Short-range surface plasmonics: Localized electron emission dynamics from a 60-nm spot on an atomically flat single-crystalline gold surface},\ }\href@noop {} {\bibfield  {journal} {\bibinfo  {journal} {Science Advances}\ }\textbf {\bibinfo {volume} {3}},\ \bibinfo {pages} {e1700721} (\bibinfo
  {year} {2017})}\BibitemShut {NoStop}%
\bibitem [{Lum()}]{Lumerical}%
  \BibitemOpen
  \href@noop {} {}\bibinfo {howpublished} {Ansys-Lumerical-FDTD, \url{https://www.ansys.com}}\BibitemShut {NoStop}%
\bibitem [{\citenamefont {Musumeci}\ \emph {et~al.}(2010)\citenamefont {Musumeci}, \citenamefont {Cultrera}, \citenamefont {Ferrario}, \citenamefont {Filippetto}, \citenamefont {Gatti}, \citenamefont {Gutierrez}, \citenamefont {Moody}, \citenamefont {Moore}, \citenamefont {Rosenzweig}, \citenamefont {Scoby} \emph {et~al.}}]{musumeci2010multiphoton}%
  \BibitemOpen
  \bibfield  {author} {\bibinfo {author} {\bibfnamefont {P.}~\bibnamefont {Musumeci}}, \bibinfo {author} {\bibfnamefont {L.}~\bibnamefont {Cultrera}}, \bibinfo {author} {\bibfnamefont {M.}~\bibnamefont {Ferrario}}, \bibinfo {author} {\bibfnamefont {D.}~\bibnamefont {Filippetto}}, \bibinfo {author} {\bibfnamefont {G.}~\bibnamefont {Gatti}}, \bibinfo {author} {\bibfnamefont {M.}~\bibnamefont {Gutierrez}}, \bibinfo {author} {\bibfnamefont {J.}~\bibnamefont {Moody}}, \bibinfo {author} {\bibfnamefont {N.}~\bibnamefont {Moore}}, \bibinfo {author} {\bibfnamefont {J.}~\bibnamefont {Rosenzweig}}, \bibinfo {author} {\bibfnamefont {C.}~\bibnamefont {Scoby}}, \emph {et~al.},\ }\bibfield  {title} {\bibinfo {title} {Multiphoton photoemission from a copper cathode illuminated by ultrashort laser pulses in an rf photoinjector},\ }\href@noop {} {\bibfield  {journal} {\bibinfo  {journal} {Physical review letters}\ }\textbf {\bibinfo {volume} {104}},\ \bibinfo {pages} {084801} (\bibinfo {year} {2010})}\BibitemShut {NoStop}%
\bibitem [{\citenamefont {Ferrini}\ \emph {et~al.}(2009)\citenamefont {Ferrini}, \citenamefont {Banfi}, \citenamefont {Giannetti},\ and\ \citenamefont {Parmigiani}}]{ferrini2009non}%
  \BibitemOpen
  \bibfield  {author} {\bibinfo {author} {\bibfnamefont {G.}~\bibnamefont {Ferrini}}, \bibinfo {author} {\bibfnamefont {F.}~\bibnamefont {Banfi}}, \bibinfo {author} {\bibfnamefont {C.}~\bibnamefont {Giannetti}},\ and\ \bibinfo {author} {\bibfnamefont {F.}~\bibnamefont {Parmigiani}},\ }\bibfield  {title} {\bibinfo {title} {Non-linear electron photoemission from metals with ultrashort pulses},\ }\href@noop {} {\bibfield  {journal} {\bibinfo  {journal} {Nuclear Instruments and Methods in Physics Research Section A: Accelerators, Spectrometers, Detectors and Associated Equipment}\ }\textbf {\bibinfo {volume} {601}},\ \bibinfo {pages} {123} (\bibinfo {year} {2009})}\BibitemShut {NoStop}%
\bibitem [{Foc()}]{Focus_PEEM}%
  \BibitemOpen
  \href@noop {} {}\bibinfo {howpublished} {FOCUS-IS-IEF-PEEM, \url{https://www.focus-gmbh.com}}\BibitemShut {NoStop}%
\bibitem [{\citenamefont {Pierce}\ \emph {et~al.}(2023)\citenamefont {Pierce}, \citenamefont {Durham}, \citenamefont {Riminucci}, \citenamefont {Dhuey}, \citenamefont {Bazarov}, \citenamefont {Maxson}, \citenamefont {Minor},\ and\ \citenamefont {Filippetto}}]{pierce2023experimental}%
  \BibitemOpen
  \bibfield  {author} {\bibinfo {author} {\bibfnamefont {C.~M.}\ \bibnamefont {Pierce}}, \bibinfo {author} {\bibfnamefont {D.~B.}\ \bibnamefont {Durham}}, \bibinfo {author} {\bibfnamefont {F.}~\bibnamefont {Riminucci}}, \bibinfo {author} {\bibfnamefont {S.}~\bibnamefont {Dhuey}}, \bibinfo {author} {\bibfnamefont {I.}~\bibnamefont {Bazarov}}, \bibinfo {author} {\bibfnamefont {J.}~\bibnamefont {Maxson}}, \bibinfo {author} {\bibfnamefont {A.~M.}\ \bibnamefont {Minor}},\ and\ \bibinfo {author} {\bibfnamefont {D.}~\bibnamefont {Filippetto}},\ }\bibfield  {title} {\bibinfo {title} {Experimental characterization of photoemission from plasmonic nanogroove arrays},\ }\href@noop {} {\bibfield  {journal} {\bibinfo  {journal} {Physical Review Applied}\ }\textbf {\bibinfo {volume} {19}},\ \bibinfo {pages} {034034} (\bibinfo {year} {2023})}\BibitemShut {NoStop}%
\bibitem [{\citenamefont {Dreher}\ \emph {et~al.}(2023)\citenamefont {Dreher}, \citenamefont {Janoschka}, \citenamefont {Frank}, \citenamefont {Giessen},\ and\ \citenamefont {Meyer~zu Heringdorf}}]{dreher2023focused}%
  \BibitemOpen
  \bibfield  {author} {\bibinfo {author} {\bibfnamefont {P.}~\bibnamefont {Dreher}}, \bibinfo {author} {\bibfnamefont {D.}~\bibnamefont {Janoschka}}, \bibinfo {author} {\bibfnamefont {B.}~\bibnamefont {Frank}}, \bibinfo {author} {\bibfnamefont {H.}~\bibnamefont {Giessen}},\ and\ \bibinfo {author} {\bibfnamefont {F.-J.}\ \bibnamefont {Meyer~zu Heringdorf}},\ }\bibfield  {title} {\bibinfo {title} {Focused surface plasmon polaritons coherently couple to electronic states in above-threshold electron emission},\ }\href@noop {} {\bibfield  {journal} {\bibinfo  {journal} {Communications Physics}\ }\textbf {\bibinfo {volume} {6}},\ \bibinfo {pages} {15} (\bibinfo {year} {2023})}\BibitemShut {NoStop}%
\bibitem [{\citenamefont {Ramchandani}(1970)}]{ramchandani1970energy}%
  \BibitemOpen
  \bibfield  {author} {\bibinfo {author} {\bibfnamefont {M.}~\bibnamefont {Ramchandani}},\ }\bibfield  {title} {\bibinfo {title} {Energy band structure of gold},\ }\href@noop {} {\bibfield  {journal} {\bibinfo  {journal} {Journal of Physics C: Solid State Physics}\ }\textbf {\bibinfo {volume} {3}},\ \bibinfo {pages} {S1} (\bibinfo {year} {1970})}\BibitemShut {NoStop}%
\bibitem [{\citenamefont {Rangel}\ \emph {et~al.}(2012)\citenamefont {Rangel}, \citenamefont {Kecik}, \citenamefont {Trevisanutto}, \citenamefont {Rignanese}, \citenamefont {Van~Swygenhoven},\ and\ \citenamefont {Olevano}}]{rangel2012band}%
  \BibitemOpen
  \bibfield  {author} {\bibinfo {author} {\bibfnamefont {T.}~\bibnamefont {Rangel}}, \bibinfo {author} {\bibfnamefont {D.}~\bibnamefont {Kecik}}, \bibinfo {author} {\bibfnamefont {P.}~\bibnamefont {Trevisanutto}}, \bibinfo {author} {\bibfnamefont {G.-M.}\ \bibnamefont {Rignanese}}, \bibinfo {author} {\bibfnamefont {H.}~\bibnamefont {Van~Swygenhoven}},\ and\ \bibinfo {author} {\bibfnamefont {V.}~\bibnamefont {Olevano}},\ }\bibfield  {title} {\bibinfo {title} {Band structure of gold from many-body perturbation theory},\ }\href@noop {} {\bibfield  {journal} {\bibinfo  {journal} {Physical Review B}\ }\textbf {\bibinfo {volume} {86}},\ \bibinfo {pages} {125125} (\bibinfo {year} {2012})}\BibitemShut {NoStop}%
\bibitem [{\citenamefont {Kupratakuln}(1970)}]{kupratakuln1970relativistic}%
  \BibitemOpen
  \bibfield  {author} {\bibinfo {author} {\bibfnamefont {S.}~\bibnamefont {Kupratakuln}},\ }\bibfield  {title} {\bibinfo {title} {Relativistic electron band structure of gold},\ }\href@noop {} {\bibfield  {journal} {\bibinfo  {journal} {Journal of Physics C: Solid State Physics}\ }\textbf {\bibinfo {volume} {3}},\ \bibinfo {pages} {S109} (\bibinfo {year} {1970})}\BibitemShut {NoStop}%
\bibitem [{\citenamefont {Bormann}\ \emph {et~al.}(2010)\citenamefont {Bormann}, \citenamefont {Gulde}, \citenamefont {Weismann}, \citenamefont {Yalunin},\ and\ \citenamefont {Ropers}}]{bormann2010tip}%
  \BibitemOpen
  \bibfield  {author} {\bibinfo {author} {\bibfnamefont {R.}~\bibnamefont {Bormann}}, \bibinfo {author} {\bibfnamefont {M.}~\bibnamefont {Gulde}}, \bibinfo {author} {\bibfnamefont {A.}~\bibnamefont {Weismann}}, \bibinfo {author} {\bibfnamefont {S.~V.}\ \bibnamefont {Yalunin}},\ and\ \bibinfo {author} {\bibfnamefont {C.}~\bibnamefont {Ropers}},\ }\bibfield  {title} {\bibinfo {title} {Tip-enhanced strong-field photoemission},\ }\href@noop {} {\bibfield  {journal} {\bibinfo  {journal} {Physical review letters}\ }\textbf {\bibinfo {volume} {105}},\ \bibinfo {pages} {147601} (\bibinfo {year} {2010})}\BibitemShut {NoStop}%
\bibitem [{\citenamefont {Bazarov}\ \emph {et~al.}(2009)\citenamefont {Bazarov}, \citenamefont {Dunham},\ and\ \citenamefont {Sinclair}}]{bazarov2009maximum}%
  \BibitemOpen
  \bibfield  {author} {\bibinfo {author} {\bibfnamefont {I.~V.}\ \bibnamefont {Bazarov}}, \bibinfo {author} {\bibfnamefont {B.~M.}\ \bibnamefont {Dunham}},\ and\ \bibinfo {author} {\bibfnamefont {C.~K.}\ \bibnamefont {Sinclair}},\ }\bibfield  {title} {\bibinfo {title} {Maximum achievable beam brightness from photoinjectors},\ }\href@noop {} {\bibfield  {journal} {\bibinfo  {journal} {Physical review letters}\ }\textbf {\bibinfo {volume} {102}},\ \bibinfo {pages} {104801} (\bibinfo {year} {2009})}\BibitemShut {NoStop}%
\bibitem [{\citenamefont {Dowell}(2019)}]{dowell2019topological}%
  \BibitemOpen
  \bibfield  {author} {\bibinfo {author} {\bibfnamefont {D.~H.}\ \bibnamefont {Dowell}},\ }\bibfield  {title} {\bibinfo {title} {Topological cathodes: Controlling the space charge limit of electron emission using metamaterials},\ }\href@noop {} {\bibfield  {journal} {\bibinfo  {journal} {Physical Review Accelerators and Beams}\ }\textbf {\bibinfo {volume} {22}},\ \bibinfo {pages} {084201} (\bibinfo {year} {2019})}\BibitemShut {NoStop}%
\bibitem [{\citenamefont {Meier}\ \emph {et~al.}(2023)\citenamefont {Meier}, \citenamefont {Heimerl},\ and\ \citenamefont {Hommelhoff}}]{meier2023few}%
  \BibitemOpen
  \bibfield  {author} {\bibinfo {author} {\bibfnamefont {S.}~\bibnamefont {Meier}}, \bibinfo {author} {\bibfnamefont {J.}~\bibnamefont {Heimerl}},\ and\ \bibinfo {author} {\bibfnamefont {P.}~\bibnamefont {Hommelhoff}},\ }\bibfield  {title} {\bibinfo {title} {Few-electron correlations after ultrafast photoemission from nanometric needle tips},\ }\href@noop {} {\bibfield  {journal} {\bibinfo  {journal} {Nature Physics}\ }\textbf {\bibinfo {volume} {19}},\ \bibinfo {pages} {1402} (\bibinfo {year} {2023})}\BibitemShut {NoStop}%
\bibitem [{\citenamefont {Faillace}(2014)}]{faillace2014recent}%
  \BibitemOpen
  \bibfield  {author} {\bibinfo {author} {\bibfnamefont {L.}~\bibnamefont {Faillace}},\ }\bibfield  {title} {\bibinfo {title} {Recent advancements of rf guns},\ }\href@noop {} {\bibfield  {journal} {\bibinfo  {journal} {Physics Procedia}\ }\textbf {\bibinfo {volume} {52}},\ \bibinfo {pages} {100} (\bibinfo {year} {2014})}\BibitemShut {NoStop}%
\bibitem [{\citenamefont {Kr{\"u}ger}\ \emph {et~al.}(2007)\citenamefont {Kr{\"u}ger}, \citenamefont {Dufft}, \citenamefont {Koter},\ and\ \citenamefont {Hertwig}}]{kruger2007femtosecond}%
  \BibitemOpen
  \bibfield  {author} {\bibinfo {author} {\bibfnamefont {J.}~\bibnamefont {Kr{\"u}ger}}, \bibinfo {author} {\bibfnamefont {D.}~\bibnamefont {Dufft}}, \bibinfo {author} {\bibfnamefont {R.}~\bibnamefont {Koter}},\ and\ \bibinfo {author} {\bibfnamefont {A.}~\bibnamefont {Hertwig}},\ }\bibfield  {title} {\bibinfo {title} {Femtosecond laser-induced damage of gold films},\ }\href@noop {} {\bibfield  {journal} {\bibinfo  {journal} {Applied surface science}\ }\textbf {\bibinfo {volume} {253}},\ \bibinfo {pages} {7815} (\bibinfo {year} {2007})}\BibitemShut {NoStop}%
\bibitem [{\citenamefont {England}\ \emph {et~al.}(2014)\citenamefont {England}, \citenamefont {Noble}, \citenamefont {Bane}, \citenamefont {Dowell}, \citenamefont {Ng}, \citenamefont {Spencer}, \citenamefont {Tantawi}, \citenamefont {Wu}, \citenamefont {Byer}, \citenamefont {Peralta} \emph {et~al.}}]{england2014dielectric}%
  \BibitemOpen
  \bibfield  {author} {\bibinfo {author} {\bibfnamefont {R.~J.}\ \bibnamefont {England}}, \bibinfo {author} {\bibfnamefont {R.~J.}\ \bibnamefont {Noble}}, \bibinfo {author} {\bibfnamefont {K.}~\bibnamefont {Bane}}, \bibinfo {author} {\bibfnamefont {D.~H.}\ \bibnamefont {Dowell}}, \bibinfo {author} {\bibfnamefont {C.-K.}\ \bibnamefont {Ng}}, \bibinfo {author} {\bibfnamefont {J.~E.}\ \bibnamefont {Spencer}}, \bibinfo {author} {\bibfnamefont {S.}~\bibnamefont {Tantawi}}, \bibinfo {author} {\bibfnamefont {Z.}~\bibnamefont {Wu}}, \bibinfo {author} {\bibfnamefont {R.~L.}\ \bibnamefont {Byer}}, \bibinfo {author} {\bibfnamefont {E.}~\bibnamefont {Peralta}}, \emph {et~al.},\ }\bibfield  {title} {\bibinfo {title} {Dielectric laser accelerators},\ }\href@noop {} {\bibfield  {journal} {\bibinfo  {journal} {Reviews of Modern Physics}\ }\textbf {\bibinfo {volume} {86}},\ \bibinfo {pages} {1337} (\bibinfo {year} {2014})}\BibitemShut {NoStop}%
\bibitem [{\citenamefont {Simakov}\ \emph {et~al.}(2017)\citenamefont {Simakov}, \citenamefont {Andrews}, \citenamefont {Herman}, \citenamefont {Hubbard},\ and\ \citenamefont {Weis}}]{simakov2017diamond}%
  \BibitemOpen
  \bibfield  {author} {\bibinfo {author} {\bibfnamefont {E.~I.}\ \bibnamefont {Simakov}}, \bibinfo {author} {\bibfnamefont {H.~L.}\ \bibnamefont {Andrews}}, \bibinfo {author} {\bibfnamefont {M.~J.}\ \bibnamefont {Herman}}, \bibinfo {author} {\bibfnamefont {K.~M.}\ \bibnamefont {Hubbard}},\ and\ \bibinfo {author} {\bibfnamefont {E.}~\bibnamefont {Weis}},\ }\bibfield  {title} {\bibinfo {title} {Diamond field emitter array cathodes and possibilities of employing additive manufacturing for dielectric laser accelerating structures},\ }in\ \href@noop {} {\emph {\bibinfo {booktitle} {AIP Conference Proceedings}}},\ Vol.\ \bibinfo {volume} {1812}\ (\bibinfo {organization} {AIP Publishing},\ \bibinfo {year} {2017})\BibitemShut {NoStop}%
\bibitem [{\citenamefont {Dai}\ \emph {et~al.}(2021)\citenamefont {Dai}, \citenamefont {Zhou}, \citenamefont {Ghosh}, \citenamefont {Yang}, \citenamefont {Huang},\ and\ \citenamefont {Petek}}]{dai2021ultrafast}%
  \BibitemOpen
  \bibfield  {author} {\bibinfo {author} {\bibfnamefont {Y.}~\bibnamefont {Dai}}, \bibinfo {author} {\bibfnamefont {Z.}~\bibnamefont {Zhou}}, \bibinfo {author} {\bibfnamefont {A.}~\bibnamefont {Ghosh}}, \bibinfo {author} {\bibfnamefont {S.}~\bibnamefont {Yang}}, \bibinfo {author} {\bibfnamefont {C.-B.}\ \bibnamefont {Huang}},\ and\ \bibinfo {author} {\bibfnamefont {H.}~\bibnamefont {Petek}},\ }\bibfield  {title} {\bibinfo {title} {Ultrafast nanofemto photoemission electron microscopy of vectorial plasmonic fields},\ }\href@noop {} {\bibfield  {journal} {\bibinfo  {journal} {MRS Bulletin}\ }\textbf {\bibinfo {volume} {46}},\ \bibinfo {pages} {738} (\bibinfo {year} {2021})}\BibitemShut {NoStop}%
\bibitem [{\citenamefont {Graves}\ \emph {et~al.}(2020)\citenamefont {Graves}, \citenamefont {Fromme}, \citenamefont {Holl}, \citenamefont {Malin}, \citenamefont {Messerschmidt}, \citenamefont {Nanni}, \citenamefont {Sandhu}, \citenamefont {Tantawi}, \citenamefont {Tilton},\ and\ \citenamefont {Spence}}]{graves2020asu}%
  \BibitemOpen
  \bibfield  {author} {\bibinfo {author} {\bibfnamefont {W.}~\bibnamefont {Graves}}, \bibinfo {author} {\bibfnamefont {P.}~\bibnamefont {Fromme}}, \bibinfo {author} {\bibfnamefont {M.}~\bibnamefont {Holl}}, \bibinfo {author} {\bibfnamefont {L.}~\bibnamefont {Malin}}, \bibinfo {author} {\bibfnamefont {M.}~\bibnamefont {Messerschmidt}}, \bibinfo {author} {\bibfnamefont {E.}~\bibnamefont {Nanni}}, \bibinfo {author} {\bibfnamefont {A.}~\bibnamefont {Sandhu}}, \bibinfo {author} {\bibfnamefont {S.}~\bibnamefont {Tantawi}}, \bibinfo {author} {\bibfnamefont {S.}~\bibnamefont {Tilton}},\ and\ \bibinfo {author} {\bibfnamefont {J.}~\bibnamefont {Spence}},\ }\bibfield  {title} {\bibinfo {title} {The asu compact xfel project},\ }\href@noop {} {\bibfield  {journal} {\bibinfo  {journal} {Bulletin of the American Physical Society}\ }\textbf {\bibinfo {volume} {65}} (\bibinfo {year} {2020})}\BibitemShut {NoStop}%
\bibitem [{\citenamefont {Conde}\ \emph {et~al.}(2017)\citenamefont {Conde}, \citenamefont {Antipov}, \citenamefont {Doran}, \citenamefont {Gai}, \citenamefont {Gao}, \citenamefont {Ha}, \citenamefont {Jing}, \citenamefont {Liu}, \citenamefont {Neveu}, \citenamefont {Power} \emph {et~al.}}]{conde2017research}%
  \BibitemOpen
  \bibfield  {author} {\bibinfo {author} {\bibfnamefont {M.}~\bibnamefont {Conde}}, \bibinfo {author} {\bibfnamefont {S.}~\bibnamefont {Antipov}}, \bibinfo {author} {\bibfnamefont {D.}~\bibnamefont {Doran}}, \bibinfo {author} {\bibfnamefont {W.}~\bibnamefont {Gai}}, \bibinfo {author} {\bibfnamefont {Q.}~\bibnamefont {Gao}}, \bibinfo {author} {\bibfnamefont {G.}~\bibnamefont {Ha}}, \bibinfo {author} {\bibfnamefont {C.}~\bibnamefont {Jing}}, \bibinfo {author} {\bibfnamefont {W.}~\bibnamefont {Liu}}, \bibinfo {author} {\bibfnamefont {N.}~\bibnamefont {Neveu}}, \bibinfo {author} {\bibfnamefont {J.}~\bibnamefont {Power}}, \emph {et~al.},\ }\bibfield  {title} {\bibinfo {title} {Research program and recent results at the argonne wakefield accelerator facility (awa)},\ }\href@noop {} {\bibfield  {journal} {\bibinfo  {journal} {Proc. IPAC’17}\ ,\ \bibinfo {pages} {2885}} (\bibinfo {year} {2017})}\BibitemShut {NoStop}%
\end{thebibliography}%

\end{document}